\documentclass[12pt]{article}
\usepackage{amsthm, bm, graphicx, hyperref, mathrsfs, bbm}
\usepackage{amsfonts}
\usepackage{amsmath}
\usepackage{amssymb}
\usepackage{dsfont}
\usepackage{graphicx}
\usepackage{color}
\usepackage[all, knot]{xy}
\usepackage{tikz}
\usepackage{braket}
\usepackage{epstopdf}
\usepackage[footnotesize]{caption}
\usepackage{bigfoot}
\usepackage{amsthm}
\usepackage{enumitem}
\usepackage{mathrsfs}
\usepackage{mathtools}     
\usepackage{subeqnarray}         
\usepackage{cases}               
\usepackage{color}
\usepackage{subfigure}
\usepackage{cite}               
\usepackage{hyperref}            
\usepackage{multirow,makecell}   
\usepackage{textcomp}
\usepackage{wasysym}
\usepackage{url}
\usepackage[margin=3cm]{geometry}
\usepackage{geometry}
\usepackage{amsfonts}
\usepackage{graphicx}
\usepackage{float}
\usepackage{amsmath}
\usepackage{hyperref}
\usepackage{tensor}
\usepackage{subfigure}
\usepackage{cases}
\usepackage{appendix}
\usepackage{multirow}
\usepackage{color}
\allowdisplaybreaks[3]

\newcommand*{\dif}{\mathop{}\!\mathrm{d}}
\DeclareNewFootnote{default} 
\DeclareNewFootnote{B} 

\begin{document}
\begin{titlepage}
\vspace{0.5cm}
\begin{center}
{\Large\bf{Gravitational formulation of stress-tensor deformed field theories}}
\vskip 1.5cm
{\large{Yunfei Xie$^{a,}$\footnote{jieyf22@mails.jlu.edu.cn}, Yun-Ze Li$^{a,}$\footnote{lyz21@mails.jlu.edu.cn}, Lixin Xu$^{b,}$\footnote{lxxu@dlut.edu.cn (Corresponding author)  }, Song He$^{c,a}$\footnote{hesong@nbu.edu.cn (Corresponding author)}}}
\vskip 2.5em
{\normalsize\it $^a$Center for Theoretical Physics and College of Physics, Jilin University, Changchun 130012, People's Republic of China\\
$^b${Institute of Theoretical Physics, School of Physics, Dalian University of Technology, Dalian, 116024, China}\\
$^c${Institute of Fundamental Physics and Quantum Technology \& School of Physical Science and Technology, Ningbo University, Ningbo, Zhejiang 315211, China}}

\end{center}
\begin{abstract}
Stress-tensor deformations suggest a geometric origin of emergent gravity but are typically non-local for $d>2$. We couple a seed QFT to Einstein gravity with deformation parameter $\lambda$ and evaluate the gravitational path integral at the metric saddle. Around a fixed reference background, the leading deformation is universal: a bilocal term quadratic in the stress tensor with kernel set by the graviton Green’s function, plus a systematic higher-order expansion. Expressed on the saddle-point (deformed) metric, the flow becomes local. We then provide two constructive completions on deformed backgrounds—Palatini $f(R)$ gravity and an eigenvalue method for general Ricci-based theories—and apply them to scalar generalized Nambu–Goto and $\det T$ deformations (arbitrary $d$), two-dimensional multi-scalar ModMax and Born–Infeld models, and four-dimensional root-$T\bar T$ and $T\bar T$ flows of Maxwell theory yielding ModMax and Born–Infeld electrodynamics. In free field theory, an off-shell analysis further shows that the leading quantum correction generates an Einstein–Hilbert term with controlled higher-derivative terms.
\end{abstract}

\end{titlepage}
\newpage
\tableofcontents
\section{Introduction}
Emergent gravity has long been a notable and significant topic in theoretical physics \cite{Sakharov:1967nyk, Visser:2002ew}. The central idea is to interpret spacetime geometry and its dynamics as emergent phenomena arising from quantum field theory. One of the most successful realizations of this idea is the holographic duality, particularly the AdS/CFT correspondence \cite{Maldacena:1997re, Gubser:1998bc, Witten:1998qj}, which provides a way to reconstruct spacetime from boundary QFT data under specific boundary conditions. Other non-holographic realizations of emergent gravity have also been explored. In particular, gravity has been interpreted as a consequence of spontaneous symmetry breaking in \cite{Zee:1978wi, Adler:1980pg, Adler:1980bx, Zee:1980sj, Adler:1982ri}, and as a thermodynamically emergent phenomenon in \cite{Jacobson:1995ab, Padmanabhan:2009vy, Verlinde:2010hp, Padmanabhan:2010xh}. It is worth noting that any emergent gravity scenario must respect the Weinberg–Witten theorem \cite{Weinberg:1980kq, Jenkins:2009un}, which forbids a massless spin-2 particle from emerging within a local quantum field theory that possesses a Lorentz-covariant stress tensor.

A parallel line of inquiry that has gained significant attention in recent years concerns stress-tensor deformations of quantum field theories. The most representative example is the $T\bar{T}$ deformation \cite{Zamolodchikov:2004ce, Smirnov:2016lqw, Cavaglia:2016oda}. As an irrelevant deformation in two dimensions, it exhibits remarkable properties such as integrability and non-locality, and admits a well-defined holographic dual \cite{McGough:2016lol, Guica:2019nzm}. These features have led to intriguing links between the $T\bar{T}$ deformation and a wide range of topics, including string theory \cite{Giveon:2017nie, Giveon:2017myj, Baggio:2018gct, Dei:2018jyj, Chakraborty:2019mdf, Callebaut:2019omt, Apolo:2019zai}, integrable systems \cite{Cavaglia:2016oda, Conti:2018jho, Pozsgay:2019ekd, Marchetto:2019yyt, Cardy:2020olv, Jiang:2020nnb, Ferko:2024ali} and 2d quantum gravity \cite{Dubovsky:2017cnj, Dubovsky:2018bmo, Tolley:2019nmm, Okumura:2020dzb, Iliesiu:2020zld, Ebert:2022ehb, Bhattacharyya:2023gvg}. Another significant example is the root-$T\bar{T}$ deformation \cite{Babaei-Aghbolagh:2022uij, Conti:2022egv, Ferko:2022cix, Babaei-Aghbolagh:2022leo}, which commutes with the $T\bar{T}$ deformation and has also been shown to preserve integrability for a large class of classical integrable models \cite{Borsato:2022tmu}. The holographic dictionary for root-$T\bar{T}$ deformation has been studied in \cite{Ebert:2023tih}. Generalized (root-)$T\bar{T}$-like deformations in two and higher dimensions have likewise been extensively investigated in the literature \cite{Taylor:2018xcy, Bonelli:2018kik, Conti:2022egv, Morone:2024ffm, Babaei-Aghbolagh:2024hti, Tsolakidis:2024wut, Babaei-Aghbolagh:2024uqp, Babaei-Aghbolagh:2025lko, Brizio:2026ynw}.

A noteworthy aspect of these deformations driven by the stress tensor is that they admit rich geometric interpretations. For instance, the $T\bar{T}$ deformation of a field theory has been proposed to admit several equivalent formulations: as a dynamical coordinate transformation \cite{Dubovsky:2017cnj, Conti:2018tca, Caputa:2020lpa}, as coupling the undeformed theory to a flat space Jackiw-Teitelboim-like gravity \cite{Dubovsky:2017cnj, Dubovsky:2018bmo}, or equivalently, to a ghost-free massive gravity \cite{Tolley:2019nmm, Nix:2025plr} or a random geometry \cite{Cardy:2018sdv}. For a one-parameter family of $T\bar{T}$-like deformations in $d\geq2$, a metric approach was proposed in \cite{Conti:2022egv}, which establishes a dynamical equivalence between the deformed theory on the original metric and the undeformed theory on a deformed metric. It was further shown in \cite{Morone:2024ffm} that a deformed theory coupled to the Einstein gravity in the Palatini formalism is dynamically equivalent to the undeformed theory coupled to a modified gravity, such as $f(\mathcal{R})$ gravity or Ricci-based gravity. In that case, the $T\bar{T}$-like dressing can be reinterpreted as a process of gravitational reframing. The geometric realization of more general stress-tensor deformations has been investigated in \cite{Ran:2024vgl, Morone:2024sdg, Babaei-Aghbolagh:2024hti}. The relation between the flow equations satisfied by the deformed metric and Ricci-Bourguignon flows \cite{hamilton1982three, catino2017ricci} has also been discussed in \cite{Brizio:2024arr, Morone:2024sdg}. 

These formulations connect stress-tensor deformations to changes in spacetime geometry, providing clues to a new pathway toward emergent gravity. A natural idea is to investigate stress-tensor deformations that directly correspond to gravitational theories with dynamical degrees of freedom, such as Einstein gravity, rather than as modifications to an existing gravity. In \cite{Kawamoto:2025oko}, a non-local $T\bar{T}$ deformation was introduced to create traversable wormholes between two CFTs in AdS/CFT, which is defined by inserting a non-local operator $F(-\square)$ between the two stress tensors in the $T\bar{T}$ operator. Following \cite{Cardy:2018sdv, Conti:2022egv}, such a deformation can be realized via a Hubbard-Stratonovich transformation, in which an auxiliary metric is introduced to reproduce the deformation at its saddle point. The presence of $F(-\square)$ then renders the auxiliary metric dynamical. Therefore, it inspires us to investigate the correspondence between stress-tensor deformations and dynamical gravities.

In \cite{Li:2025lpa}, we presented a general framework to realize stress-tensor deformations at the metric saddle of gravitational theories. Within this framework, we derived the non-local stress-tensor deformation that corresponds to linearized Einstein gravity. By rewriting the deformed action as an effective deformed action on the saddle-point metric, one can reformulate a non-local deformation on the reference metric into a local deformation on the saddle-point metric. This thereby establishes a connection between the conventional local stress-tensor deformations and the on-shell dynamics of gravitational theories. Furthermore, when the non-local deformation induced by Einstein gravity is placed on a dynamical background, the one-loop contribution generated via the heat kernel expansion reproduces Einstein gravity, thereby serving as a check of classical-quantum consistency and offering a possible pathway toward emergent gravity\footnote{There are also alternative approaches that relate stress-tensor deformations to emergent gravity, for example, by introducing a hidden sector \cite{Betzios:2020sro} or through holographic renormalization \cite{Adami:2025pqr}.}. The nonlocal structure lies outside the locality assumptions of the Weinberg–Witten theorem. Taken together, the framework provides a semiclassical correspondence relating deformed field theories to gravity. In this paper, we will elaborate on the details. Additionally, as important examples of effective deformed actions generated from gravitational actions, we construct the gravitational actions corresponding to scalar field theories and 4d Maxwell's theory under both $T\bar{T}$ and root-$T\bar{T}$ deformations. Our constructions proceed along two paths: one by directly constructing the corresponding Palatini $f(\mathcal{R})$ gravity, and the other by employing the eigenvalue method to formulate Ricci-based gravity in terms of eigenvalues.

The rest of the paper is organized as follows.
In section~\ref{sec:framework}, we formulate the general proposal for geometric realizations of stress-tensor deformations and introduce the effective deformed action on the deformed background.
Section~\ref{section nonlocal} focuses on the Einstein gravity, deriving the non-local stress-tensor deformation induced by integrating out metric fluctuations around a fixed reference geometry.
In sections~\ref{sec:scalars} and~\ref{sec:gauge} we apply the framework to scalar theories and Maxwell's theory, and construct corresponding gravitational actions for representative $T\bar T$-like flows via Palatini $f(\mathcal{R})$ gravity and eigenvalue methods. Section~\ref{sec:quantum} studies the non-local deformation on a dynamical background and demonstrates classical-quantum consistency through the emergence of an induced Einstein-Hilbert term at one loop in the massive scalar case. We conclude in section~\ref{sec:conclusion} with a discussion of implications and open problems, including unitarity and non-perturbative aspects of the resulting non-local deformations.

\section{Gravitational description of stress-tensor deformations}\label{sec:framework}
In this section, we present our general proposal for the geometric realization of stress-tensor deformations, which allows us to investigate non-local stress-tensor deformations on a fixed background and generic stress-tensor deformations on a deformed background. We also provide a practical approach to construct gravitational theories within the Palatini formalism.

\subsection{General proposal}\label{subsection general proposal}
We start with a gravitational theory $S^{(\lambda)}_{\mathrm{grav}}[g,\psi]$ in an arbitrary dimension, where $\psi$ collectively denotes the matter fields. The action $S^{(\lambda)}_{\mathrm{grav}}$ depends on a continuous parameter $\lambda$, which will serve as the deformation parameter in the stress-tensor deformed theory. In general, the matter fields can be non-minimally coupled with the curvature invariants. The partition function is defined as
\begin{equation}
	\mathcal{Z}_{\mathrm{grav}}^{(\lambda)}=\int\mathcal{D}\psi\mathcal{D}g\,e^{-S^{(\lambda)}_{\mathrm{grav}}[g,\psi]}.
\end{equation}
Formally, the path integral of $g_{\mu\nu}$ can be evaluated in the saddle-point approximation,
\begin{align}
	\mathcal{Z}_{\mathrm{grav}}^{(\lambda)}=\int\mathcal{D}\psi\sum_{g^*}\mathcal{Z}_{\mathrm{loops}}^{(\lambda)}[g^*]e^{-S^{(\lambda)}_{\mathrm{grav}}[g,\psi]},\label{saddle-point approximation}
\end{align}
where the saddle-point metric $g^*$ satisfies
\begin{equation}
	\frac{\delta}{\delta g_{\mu\nu}}S^{(\lambda)}_{\mathrm{grav}}[g,\psi]\big|_{g=g^*}=0\label{eom of g}
\end{equation}
and $\mathcal{Z}_{\mathrm{loops}}^{(\lambda)}[g^*]$ denotes the quantum corrections around each saddle point $g^*$, which are irrelevant at the classical level we consider. For our purpose, we classify the saddle points by their $\lambda\to 0$ limits $\hat{\gamma}$. Eq. (\ref{saddle-point approximation}) can be then written as
\begin{align}
	\mathcal{Z}_{\mathrm{grav}}^{(\lambda)}=\int\mathcal{D}\psi\sum_{\hat{\gamma}}\left(\sum_{\alpha}\mathcal{Z}_{\mathrm{loops}}^{(\lambda)}[g_\alpha^*]e^{-S_{\alpha}^{(\lambda)}[\hat{\gamma},\psi]}\right),
\end{align}
where $\{g_\alpha^*\}$ denotes the set of all saddle points satisfying $\lim_{\lambda\to0}g_\alpha^*=\hat{\gamma}$. For a certain $g_\alpha^*$, we define the classical deformed field theory action as
\begin{equation}
	S_{\alpha}^{(\lambda)}[\hat{\gamma},\psi]\equiv S^{(\lambda)}_{\mathrm{grav}}[g_\alpha^*,\psi].\label{definition of deformed action}
\end{equation}\par
As we will see in the section \ref{section nonlocal}, when the matter fields are minimally coupled with the gravity sector, the deformed action $S_{\alpha}^{(\lambda)}[\hat{\gamma},\psi]$ takes the form of the seed theory after a non-local stress-tensor deformation on the fixed background $\hat{\gamma}$, which can be obtained perturbatively in principle.\par
However, obtaining a compact expression for the deformed action is generally intractable, as the metric saddle point $g_{\alpha}^{*}(\lambda)$ is hard to determine non-perturbatively. On the other hand, it is tempting to investigate those familiar local deformations like $T\bar{T}$ and root-$T\bar{T}$ deformations within this framework. Therefore, we propose an alternative approach, that is, to introduce an effective deformed action (EDA) $S_{\mathrm{EDA}}^{(\lambda)}[\gamma^{(\lambda)},\psi]$ defined on a deformed metric $\gamma^{(\lambda)}$, which incorporates the metric saddle point $g_\alpha^*$,
\begin{equation}
	S_{\mathrm{EDA}}^{(\lambda)}[\gamma^{(\lambda)},\psi]\equiv S^{(\lambda)}_\alpha[\hat{\gamma},\psi].\label{definition of effective deformed action}
\end{equation}
The deformed metric $\gamma^{(\lambda)}$ is set to be a local function of $\lambda$ and the metric saddle point $g_\alpha^{*}$. (More generally, $\gamma^{(\lambda)}$ could be a functional of $g_\alpha^*$.) With different choices of $\gamma^{(\lambda)}$, the effective deformed action functional takes different forms, producing identical results when evaluated at the saddle point $g_\alpha^*$.\par
A natural and straightforward choice is to identify $\gamma^{(\lambda)}$ with $g_\alpha^*(\lambda)$ itself. In this case, it follows from (\ref{definition of deformed action}) and (\ref{definition of effective deformed action}) that the effective deformed action $S_{\mathrm{EDA}}^{(\lambda)}[g_\alpha^*,\psi]$ is just the original gravitational action with the metric on-shell, which can be obtained by substituting the metric's equation of motion (EOM) back into the gravitational action $S_{\mathrm{grav}}^{(\lambda)}[g,\psi]$ to eliminate the curvature terms. \par
By definition, we have
\begin{align}\label{general flow equation}
	&\frac{\dif}{\dif\lambda}S^{(\lambda)}_{\mathrm{EDA}}[g_{\alpha}^*,\psi]=\frac{\dif}{\dif\lambda}S_{\alpha}^{(\lambda)}[\hat{\gamma},\psi]=\frac{\dif}{\dif\lambda}S_{\mathrm{grav}}^{(\lambda)}[g_\alpha^*,\psi]\nonumber\\
	=&\ \partial_\lambda S_{\mathrm{grav}}^{(\lambda)}[g_{\alpha}^*,\psi]+\int\dif^dx\,\frac{\dif g_{\alpha,\mu\nu}^*}{\dif\lambda}\frac{\delta}{\delta g_{\mu\nu}}S^{(\lambda)}_{\mathrm{grav}}[g,\psi]\big|_{g=g_\alpha^*}\nonumber\\
	=&\ \partial_\lambda S_{\mathrm{grav}}^{(\lambda)}[g_{\alpha}^*,\psi],
\end{align}
where the partial derivative $\partial_\lambda$ denotes derivative only with respect to the explicit dependence on $\lambda$ in $S_{\mathrm{grav}}^{(\lambda)}[g_\alpha^*,\psi]$. The EOM of $g_{\mu\nu}$ (\ref{eom of g}) is applied at the last step. This approach allows us to investigate local stress-tensor deformations on the deformed background.\par
In this paper, we present two distinct lines of investigation within this framework. In section \ref{section nonlocal}, we investigate non-local stress-tensor deformation generated by Einstein's gravity on the fixed background perturbatively, and discuss its corresponding effective deformed action on the deformed background. In subsequent sections, we start from specific stress-tensor deformations on the deformed background and construct gravitational actions that generate them. To this end, we adopt the Palatini formalism, which is introduced in the following two subsections.

\subsection{Palatini $f(\mathcal{R})$ gravity}\label{Subsection 2.2}
The simplest and most direct approach to construct a gravitational theory that describes the stress-tensor deformed field theory is to consider a (non-minimal) coupling between the Palatini Ricci scalar and the matter fields. In the Palatini formalism, the variations of the metric and the connection are independent. The Riemann curvature tensor $\mathcal{R}^{\mu}{}_{\nu\rho\sigma}$ and the Ricci curvature tensor $\mathcal{R}_{\mu\nu}$ are constructed with the independent connection $\Gamma$. We assume that the field theory contains only a single tensor field $X_{\mu\nu}$, which is independent of the metric and the connection. There exist $d$ independent invariants constructed from $X_{\mu\nu}$, which can be expressed as  
\begin{align}
X_n=g^{\nu_n\mu_1}X_{\mu_1\nu_1}g^{\nu_1\mu_2}X_{\mu_2\nu_2}...g^{\nu_{n-1}\mu_n}X_{\mu_n\nu_n}=g^{\mu\nu}(X^n)_{\mu\nu},\ \ \text{for}\ \ n=1,2,...,d.
\end{align}
The gravitational Lagrangian is a local function of these invariants and the Palatini Ricci scalar $\mathcal{R}=g^{\mu\nu}\mathcal{R}_{\mu\nu}$,
\begin{align}
    S^{(\lambda)}_{\text{grav}}[g,\Gamma,\psi]&=\int d^dx\sqrt{g}\mathcal{A}^{(\lambda)}(\mathcal{R},X_1,...,X_d).
\end{align}
Taking variations with respect to the metric and the connection, we obtain
\begin{align}
\mathcal{A}^{(\lambda)}_{\mathcal{R}}\mathcal{R}_{(\mu\nu)}+\sum_{n=1}^{d}n\mathcal{A}^{(\lambda)}_{X_n}(X^n)_{\mu\nu}-\frac{1}{2}\mathcal{A}^{(\lambda)}g_{\mu\nu}&=0,\label{non-minimal f(R) EoM 1}\\
\bar\nabla_{\sigma}\Big(\sqrt{g}\mathcal{A}^{(\lambda)}_{\mathcal{R}}g^{\mu\nu}\Big)&=0,\label{non-minimal f(R) EoM 2}
\end{align}
where $\bar\nabla$ denotes the covariant derivative with respect to $\Gamma^{\sigma}{}_{\mu\nu}$. For a certain $\mathcal{A}^{(\lambda)}$, the trace of the first equation (\ref{non-minimal f(R) EoM 1}) is an algebraic equation in $\mathcal{R}$. The solution can be formally expressed as $\mathcal{R}^*=\mathcal{R}^*(X_1,...,X_d)$. Plugging it into the gravitational action, we have
\begin{align}
    S^{(\lambda)}_{\text{EDA}}[g,\psi]&=\int d^dx\sqrt{g}\mathcal{A}^{(\lambda)}(\mathcal{R}^*,X_1,...,X_d)\notag\\
    &=\int d^dx\sqrt{g}\mathcal{B}^{(\lambda)}(X_1,...,X_d).
\end{align}
The second equation (\ref{non-minimal f(R) EoM 2}) is used to determine the deformed metric $g^{(\lambda)*}$. For $d>2$, by introducing the following conformal transformation \cite{Sotiriou:2008rp,DeFelice:2010aj},
\begin{align}
    \tilde g_{\mu\nu}&=(\mathcal{A}^{(\lambda)}_{\mathcal{R}})^{\frac{2}{d-2}}g_{\mu\nu},
\end{align}
eq.(\ref{non-minimal f(R) EoM 2}) reduces to the definition of the Levi-Civita connection of $\tilde g$. The independent connection takes the form
\begin{align}
        \Gamma^{\sigma}{}_{\mu\nu}=\frac{1}{2}\tilde{g}^{\sigma\lambda}(\partial_{\mu}\tilde{g}_{\nu\lambda}+\partial_{\nu}\tilde{g}_{\mu\lambda}-\partial_{\lambda}\tilde{g}_{\mu\nu}).
\end{align}
Therefore, the Palatini Ricci tensor $\mathcal{R}_{\mu\nu}$ is compatible with the metric $\tilde{g}$. It is related to the Ricci tensor $R_{\mu\nu}$ constructed from the metric $g$ through the following conformal transformation,
\begin{align}\label{Palatini curvature to metric curvature}
            \mathcal{R}_{\mu\nu}=R_{\mu\nu}-\frac{1}{\mathcal{A}^{(\lambda)}_{\mathcal{R}}}\Big(\nabla_{\mu}\nabla_{\nu}+\frac{1}{d-2}g_{\mu\nu}\Box\Big)\mathcal{A}^{(\lambda)}_{\mathcal{R}}+\frac{d-1}{d-2}\frac{1}{(\mathcal{A}^{(\lambda)}_{\mathcal{R}})^2}\nabla_{\mu}\mathcal{A}^{(\lambda)}_{\mathcal{R}}\nabla_{\nu}\mathcal{A}^{(\lambda)}_{\mathcal{R}}.
\end{align}
Contraction with respect to $g^{\mu\nu}$ gives
\begin{equation}
\mathcal{R} = R -\frac{1}{\mathcal{A}^{(\lambda)}_{\mathcal{R}}} \frac{2(d-1)}{d-2} \Box \mathcal{A}^{(\lambda)}_{\mathcal{R}} +\frac{d-1}{d-2}\frac{1}{(\mathcal{A}^{(\lambda)}_{\mathcal{R}})^2}\nabla_{\alpha}\mathcal{A}^{(\lambda)}_{\mathcal{R}}\nabla^{\alpha}\mathcal{A}^{(\lambda)}_{\mathcal{R}}.
\end{equation}

By combining these expressions with the first equation (\ref{non-minimal f(R) EoM 1}), we find a dynamical equation that depends only on the metric,
\begin{align}
    G_{\mu\nu}&=\frac{1}{\mathcal{A}^{(\lambda)}_{\mathcal{R}}}\Big[\nabla_{\mu}\nabla_{\nu}\mathcal{A}^{(\lambda)}_{\mathcal{R}}-g_{\mu\nu}\Box\mathcal{A}^{(\lambda)}_{\mathcal{R}}-\frac{d-1}{d-2}\frac{1}{\mathcal{A}^{(\lambda)}_{\mathcal{R}}}\Big(\nabla_{\mu}\mathcal{A}^{(\lambda)}_{\mathcal{R}}\nabla_{\nu}\mathcal{A}^{(\lambda)}_{\mathcal{R}}-\frac{1}{2}g_{\mu\nu}(\nabla\mathcal{A}^{(\lambda)}_{\mathcal{R}})^2\Big)\notag\\
    &\quad-\sum_{n=1}^dn\mathcal{A}^{(\lambda)}_{X_n}\Big((X^n)_{\mu\nu}-\frac{1}{2}g_{\mu\nu}X_n\Big)-\frac{d-2}{4}\mathcal{A}^{(\lambda)}g_{\mu\nu}\Big],\label{modified Einstein's equation for single scalar field}
\end{align}
which can be interpreted as Einstein's equation with a modified source, where $G_{\mu\nu} = R_{\mu\nu} - \frac{1}{2}R g_{\mu\nu} $ is the $d$-dimensional Einstein tensor.

For $d=2$, Eq. (\ref{non-minimal f(R) EoM 2}) forces $\partial_\sigma\mathcal{A}_\mathcal{R}^{(\lambda)}=0$ and leads to
\begin{equation}
	\bar{\nabla}_\sigma g_{\mu\nu}=\frac{1}{2}g_{\mu\nu}g^{\alpha\beta}\bar{\nabla}_\sigma g_{\alpha\beta}=\omega_\sigma g_{\mu\nu},
\end{equation}
where $\omega_\sigma\equiv\frac{1}{2}g^{\alpha\beta}\bar{\nabla}_\sigma g_{\alpha\beta}$. Consequently, $\Gamma_{\ \mu\nu}^\sigma$ becomes a Weyl connection,
\begin{align}
	\Gamma_{\ \mu\nu}^\sigma&=\frac{1}{2}g^{\sigma\rho}\left(\partial_\mu g_{\rho\nu}+\partial_\nu g_{\mu\rho}-\partial_\rho g_{\mu\nu}\right)-\frac{1}{2}\left(\delta^\sigma_\mu\omega_\nu+\delta^\sigma_\nu\omega_\mu-\omega^\sigma g_{\mu\nu}\right),\nonumber\\
	&=\tilde{\Gamma}_{\ \mu\nu}^\sigma-\frac{1}{2}\left(\delta^\sigma_\mu\omega_\nu+\delta^\sigma_\nu\omega_\mu-\omega^\sigma g_{\mu\nu}\right),
\end{align}
with $\tilde{\Gamma}_{\ \mu\nu}^\sigma$ denoting the Levi-Civita connection of $g_{\mu\nu}$. The corresponding Ricci tensor and scalar curvature are then
\begin{align}
	\mathcal{R}_{\mu\nu}&=\tilde{R}_{\mu\nu}+\frac{1}{2}g_{\mu\nu}\tilde{\nabla}_\rho\omega^\rho+\frac{1}{2}\tilde{\nabla}_\nu\omega_\mu-\frac{1}{2}\tilde{\nabla}_\mu\omega_\nu,\\
	\mathcal{R}&=\tilde{R}+\tilde{\nabla}_\rho\omega^\rho,
\end{align}
where tilded quantities are constructed from the Levi-Civita connection of $g_{\mu\nu}$. Combining these with the metric EOM (\ref{non-minimal f(R) EoM 1}) gives the dynamical equation of $g_{\mu\nu}$:
\begin{align}
	\tilde{G}_{\mu\nu}=\tilde{R}_{\mu\nu}-\frac{1}{2}\tilde{R}g_{\mu\nu}=-\frac{1}{\mathcal{A}_{\mathcal{R}}^{(\lambda)}}\left[\sum_{n=1}^2 n\mathcal{A}_{X_n}^{(\lambda)}\left((X^n)_{\mu\nu}-\frac{1}{2}g_{\mu\nu}X_n\right)\right],\label{modified Einstein's equation in 2d}
\end{align}
which is independent of the choice of $\omega_\sigma$.

\subsection{Eigenvalue method}\label{subsection Eigenvalue method}
In the Palatini formulation of $f(\mathcal{R})$ gravity, taking the trace of the EOM yields an algebraic relation for the Palatini Ricci scalar, thereby allowing its elimination from the gravitational action and yielding an effective deformed field-theory action. The applicability of this approach, however, presupposes an action that contains only the Ricci scalar. For actions with more general curvature scalars, an eigenvalue-based method offers an efficient alternative. We assume that the gravitational Lagrangian $\mathcal{A}^{(\lambda)}$ is a local function of the invariants $X_n$ and of those constructed from the Palatini Ricci curvature tensor, $\mathcal{R}_n = g^{\mu\nu}(\mathcal{R}^n)_{\mu\nu}$, respectively. Taking the variation with respect to the metric, we obtain the EOM
\begin{align}\label{Eigenvalue method matric EoM}
2\sum_{m=1}^{d}m(\mathcal{R}^m)^{\mu}{}_{\nu}\mathcal{A}^{(\lambda)}_{\mathcal{R}_m}+2\sum_{n=1}^{d}n(X^n)^{\mu}{}_{\nu}\mathcal{A}^{(\lambda)}_{X_n}-\mathcal{A}^{(\lambda)}\delta^{\mu}{}_{\nu}=0.
\end{align}
Consider matrices $X^{\mu}{}_{\nu}$ and $\mathcal{R}^{\mu}{}_{\nu}$ can be respectively diagonalized as
\begin{align}\label{diagonalization}
    X^{\mu}{}_{\nu}&=U\text{diag}(\chi_1,\chi_2,...,\chi_d)U^{-1},\notag\\
    \mathcal{R}^{\mu}{}_{\nu}&=\tilde U\text{diag}(\tilde r_1,\tilde r_2,...,\tilde r_d)\tilde U^{-1}.
\end{align}
Plugging this expression into the EOM, we find
\begin{align}
&U^{-1}\tilde U\text{diag}(2\tilde r_1\mathcal{A}^{(\lambda)}_{\tilde r_1},...,2\tilde r_d\mathcal{A}^{(\lambda)}_{\tilde r_d})(U^{-1}\tilde U)^{-1}\notag\\
&\quad\quad\quad=\text{diag}(\mathcal{A}^{(\lambda)}-2\chi_1\mathcal{A}^{(\lambda)}_{\chi_1},...,\mathcal{A}^{(\lambda)}-2\chi_d\mathcal{A}^{(\lambda)}_{\chi_d}).
\end{align}
The elements of two diagonal matrices are identical, although the ordering of these elements differs. We can relabel the eigenvalues of $\mathcal{R}^{\mu}{}_{\nu}$ as $\lbrace{r_j}\rbrace$, which satisfy the differential equations
\begin{align}\label{Eigenvalue method EoM for rj}
    2r_j\mathcal{A}^{(\lambda)}_{r_j}+2\chi_j\mathcal{A}^{(\lambda)}_{\chi_j}-\mathcal{A}^{(\lambda)}=0,\ \ \text{for }j=1,2,...,d.
\end{align}
For a given $\mathcal{A}^{(\lambda)}$, (\ref{Eigenvalue method EoM for rj}) provides a set of algebraic equations for $\lbrace{r_j}\rbrace$. In principle, the expressions $r_j=r_{j}(\chi_1,...,\chi_d)$ can be determined explicitly from them. Note that the formalism permits a redundant component, which trivially satisfies equations in (\ref{Eigenvalue method EoM for rj}) and thus leaves the expressions for $\lbrace{r_j(\chi_1,...,\chi_d)}\rbrace$ unchanged. Hence, we can decompose $\mathcal{A}^{(\lambda)}$ into two parts,
\begin{align}
\mathcal{A}^{(\lambda)}(r,\chi)=\mathcal{A}^{(\lambda)}_0(r,\chi)+\tilde{\mathcal{A}}^{(\lambda)}(r,\chi).
\end{align}
The first part $\mathcal{A}^{(\lambda)}_0$ naturally satisfies the equations in (\ref{Eigenvalue method EoM for rj}), and the general form of it is given by
\begin{align}
    \mathcal{A}^{(\lambda)}_0(r,\chi)=\Big(\prod_{j=1}^{d}r_j^{1/2}\Big)F\Big(\frac{r_1}{\chi_1},...,\frac{r_d}{\chi_d}\Big),
\end{align}
where $F$ is an arbitrary function. The second part $\tilde{\mathcal{A}}^{(\lambda)}$ determines the expressions $r_j(\chi_1,...,\chi_d)$ for $j=1,...,d$. It will be shown later that for specific deformations, such as the $T\bar T$ deformation, the deformed action usually contains a specific structure corresponding to $\tilde{\mathcal{A}}^{(\lambda)}$. Thus, only $\mathcal{A}^{(\lambda)}_0$ needs to be explicitly constructed to match the remainder of the deformed action.

\section{Non-local stress-tensor deformations on fixed background}\label{section nonlocal}
The approach developed in the preceding section allows us to investigate stress-tensor deformations induced by generic gravitational theories, particularly dynamical ones. In this section, we focus on Einstein gravity. We consider the matter seed theory minimally coupled to the Einstein-Hilbert action, with $\lambda$ as the coupling:
\begin{align}
	S_{\mathrm{grav}}^{(\lambda)}[g,\psi]&=\hat{S}[g,\psi]+\frac{1}{2\lambda}S_{\mathrm{EH}}\nonumber\\
	&=\hat{S}[g,\psi]+\frac{l^2}{2\lambda}\int\dif^dx\sqrt{g}\,R.\label{Einstein gravity}
\end{align}
Here, $l$ is a length scale to balance the scaling dimension. The saddle-point condition for the metric is the Einstein field equation\footnote{Hereafter, we will omit the subscript $\alpha$ of the saddle-point metric, since we are always working at a single saddle point.}:
\begin{equation}
	R_{\mu\nu}^*-\frac{1}{2}R^*g_{\mu\nu}^*=-\lambda l^{-2}\hat{T}_{\mu\nu}^*.\label{Einstein field equation}
\end{equation}
Following the proposal in subsection \ref{subsection general proposal}, to find the deformation generated by Einstein gravity, one needs to solve the saddle-point equation (\ref{Einstein field equation}) and substitute the solution back into (\ref{Einstein gravity}). However, it's impossible to write the solution of the general Einstein field equation in closed form directly. Therefore, we turn to the perturbation approach.

\subsection{Perturbation approach}
By solving Einstein's equation perturbatively, we can obtain the form of the deformation generated by Einstein gravity order by order. We consider the metric perturbation
\begin{equation}
	g_{\mu\nu}=\hat{\gamma}_{\mu\nu}+h_{\mu\nu}
\end{equation}
around the background metric $\hat{\gamma}_{\mu\nu}$. The full gravitational action can be formally expanded in powers of $h_{\mu\nu}$:
\begin{align}
	S&=\hat S[\hat\gamma+h,\psi]+\frac{1}{2\lambda}S_{\text{EH}}[\hat\gamma+h]\notag\\
	&=\hat S[\hat\gamma,\psi]-\frac{1}{2}\int d^dx\sqrt{\hat\gamma}h_{\mu\nu}\hat T^{\mu\nu}-\frac{1}{2}\sum_{k=1}^{\infty}\int d^dx\,h_{\mu\nu}\left(\prod_{i=1}^{k}h_{\mu_i\nu_i}\right)\frac{\partial^k}{\prod_{i=1}^k\partial\hat\gamma_{\mu_i\nu_i}}\left(\sqrt{\hat\gamma}\hat T^{\mu\nu}\right)\notag\\
	&\quad+\frac{l^{2}}{2\lambda}\sum_{k=0}^{\infty}\int d^dx\sqrt{\hat\gamma}\,\mathcal{F}^{(k)},
\end{align}
where $\mathcal{F}^{(k)}$ represents the $k$-th order terms at each order in the expansion of the Einstein-Hilbert action,
\begin{align}
	\mathcal{F}^{(k)}&=\mathcal{F}^{(k)}(\hat\gamma_{\mu\nu},h_{\mu\nu},\hat\nabla_{\rho} h_{\mu\nu},\hat\nabla_{\rho}\hat\nabla_{\sigma}h_{\mu\nu})\sim h^k,\notag\\
	[\mathcal{F}^{(k)}]&=[L]^{-2}.
\end{align}
The EOM for the metric perturbation $h_{\mu\nu}$ is
\begin{align}
	-\lambda l^{-2}&\left[\hat T^{\mu\nu}+\sum_{k=1}^{\infty}\left(\prod_{i=1}^{k}h_{\mu_i\nu_i}\right)\frac{1}{\sqrt{\hat\gamma}}\left(\frac{\partial^k}{\prod_{i=1}^k\partial\hat\gamma_{\mu_i\nu_i}}\left(\sqrt{\hat\gamma}\hat T^{\mu\nu}\right)\right.\right.\nonumber\\
	&\hspace{2.5cm}\left.\left. +\,k\frac{\partial^k}{\partial\hat\gamma_{\mu\nu}\prod_{i=2}^k\partial\hat\gamma_{\mu_i\nu_i}}\left(\sqrt{\hat\gamma}\hat T^{\mu_1\nu_1}\right)\right)\right]+\sum_{k=0}^{\infty}\tilde{\mathcal{F}}^{(k)\mu\nu}=0,\label{full EOM of h}
\end{align}
where 
\begin{align}
	\tilde{\mathcal{F}}^{(k)\mu\nu}(x)\equiv\frac{1}{\sqrt{\hat\gamma(x)}}\frac{\delta}{\delta h_{\mu\nu}(x)}\int d^dx'\sqrt{\hat\gamma}\mathcal{F}^{(k+1)}\sim h^k.
\end{align}
To solve the EOM (\ref{full EOM of h}), we introduce a second expansion by expressing the solution $h_{\mu\nu}^*$ as a power series in $\lambda$,
\begin{equation}
	h_{\mu\nu}^*=\sum_{k=-\infty}^\infty\lambda^k h_{[k]\mu\nu}^*,
\end{equation}
Clearly, when 
\begin{equation}
	\tilde{\mathcal{F}}^{(0)\mu\nu}=\frac{1}{2}\hat{R}\hat{\gamma}^{\mu\nu}-\hat{R}^{\mu\nu}=0,\label{vacuum Einstein equation}
\end{equation}
that is, when the background metric $\hat{\gamma}^{\mu\nu}$ satisfies the vacuum Einstein field equation, there exists a solution for $h_{\mu\nu}^*$ such that
\begin{align}
	h^{*}_{[k]\mu\nu}=0,\text{ for }k\leqslant 0.
\end{align}
Under this condition, we can solve (\ref{full EOM of h}) order by order in $\lambda$, and obtain the expression of the deformation at corresponding orders. Note that (\ref{vacuum Einstein equation}) indeed holds due to the $\lambda$ dependence in (\ref{Einstein field equation}).

\subsection{Leading order}\label{subsection leading order}
Following the techniques introduced in \cite{Gullu:2010em, Altas:2019qcv}, we compute the Einstein-Hilbert action up to the second order in $h_{\mu\nu}$ to obtain the leading-order contribution to the deformed action. After including the gauge fixing term and the Faddeev-Popov term\cite{Christensen:1979iy,Giombi:2008vd},
\begin{align}
	S_{\text{gauge}}&=-\frac{l^{2}}{2}\int d^dx\sqrt{\hat\gamma}\hat \gamma^{\nu\rho}\hat \nabla^{\mu}\tilde h_{\mu\nu}\hat \nabla^{\sigma}\tilde h_{\rho\sigma},\notag\\
	S_{\text{ghost}}&=-\frac{l^{2}}{2}\int d^dx\sqrt{\hat\gamma}\bar \eta_{\mu}(\hat\gamma^{\mu\nu}\hat \Box+\hat R^{\mu\nu})\eta_{\nu},
\end{align}
where
\begin{equation}
	\tilde{h}_{\mu\nu}=h_{\mu\nu}-\frac{1}{2}\hat{\gamma}_{\mu\nu}h^{\alpha}_{\alpha},
\end{equation}
and $\eta_\mu$ denotes the complex-valued vector Faddeev-Popov ghost, we can obtain the first-order term $\tilde{\mathcal{F}}^{(1)\mu\nu}$:
\begin{align}
	\tilde{\mathcal{F}}^{(1)\mu\nu}&=\frac{1}{2}\hat \Box h^{\mu\nu}-\frac{1}{4}\hat\gamma^{\mu\nu}\hat \Box h^{\alpha}_{\alpha}+\frac{1}{2}\Big[\hat R^{\mu\rho\nu\sigma}+\hat R^{\nu\rho\mu\sigma}+\hat R^{\mu\rho}\hat\gamma^{\nu\sigma}+\hat R^{\nu\rho}\hat\gamma^{\mu\sigma}\notag\\
	&\quad-\hat R^{\mu\nu}\hat\gamma^{\rho\sigma}-\hat R^{\rho\sigma}\hat\gamma^{\mu\nu}-\frac{1}{2}\hat R(2\hat\gamma^{\mu\rho}\hat\gamma^{\nu\sigma}-\hat\gamma^{\mu\nu}\hat\gamma^{\rho\sigma})\Big]h_{\rho\sigma}\nonumber\\
	&=\frac{1}{2}\hat \Box h^{\mu\nu}-\frac{1}{4}\hat\gamma^{\mu\nu}\hat \Box h^{\alpha}_{\alpha}+\frac{1}{2}\left(\hat R^{\mu\rho\nu\sigma}+\hat R^{\nu\rho\mu\sigma}\right)h_{\rho\sigma}.
\end{align}
In the last step, we have used the condition (\ref{vacuum Einstein equation}). Therefore, the EOM (\ref{full EOM of h}) reduces to
\begin{equation}
	\frac{1}{2}\hat \Box h_{[1]}^{\mu\nu}-\frac{1}{4}\hat\gamma^{\mu\nu}\hat \Box h^{\ \alpha}_{[1]\alpha}+\frac{1}{2}\left(\hat R^{\mu\rho\nu\sigma}+\hat R^{\nu\rho\mu\sigma}\right)h_{[1]\rho\sigma}-\frac{1}{l^2}\hat{T}^{\mu\nu}=0
\end{equation}
at the first order in $\lambda$, from which the saddle-point metric perturbation is found to be
\begin{align}
	h^*_{\mu\nu}(x)&=-\frac{2\lambda}{l^2} \int d^dy\sqrt{\hat\gamma(y)}G_{\mu\nu\rho\sigma}(x,y)\Big(\hat T^{\rho\sigma}(y)-\frac{1}{d-2}\hat T(y)\hat\gamma^{\rho\sigma}(y)\Big)+O(\lambda^2),\label{saddle point metric at the first order}
\end{align}
where $G_{\mu\nu\rho\sigma}(x,y)$ denotes the graviton Green's function satisfying
\begin{equation}
	\left(-\hat{\gamma}^{\mu\rho}\hat{\gamma}^{\nu\sigma}\hat{\square}-\hat{R}^{\mu\rho\nu\sigma}-\hat{R}^{\nu\rho\mu\sigma}\right)_xG_{\rho\sigma\alpha\beta}(x,y)=\frac{1}{\sqrt{\hat{\gamma}}}\delta^\mu_\alpha\delta^\nu_\beta\delta^{(d)}(x-y).\label{graviton Green's function equation}
\end{equation}
Plugging (\ref{saddle point metric at the first order}) into the gravitational action, we obtain the leading-order contribution to the deformed action:
\begin{align}
	S^{(\lambda)}[\hat\gamma,\psi]=\hat S[\hat\gamma,\psi]& +\frac{\lambda}{2l^2}\int d^dxd^dy\sqrt{\hat\gamma(x)\hat\gamma(y)}\nonumber\\
	&\times G_{\mu\nu\rho\sigma}(x,y)\hat T^{\mu\nu}(x)\left(\hat T^{\rho\sigma}-\frac{1}{d-2}\hat T^{\alpha}_{\alpha}\hat\gamma^{\rho\sigma}\right)(y)+O(\lambda^2).\label{first-order deformed action}
\end{align}
Specifically, when the background metric is flat, $\hat{\gamma}_{\mu\nu}=\eta_{\mu\nu}$, the Green's function becomes
\begin{equation}
	G_{\mu\nu\rho\sigma}(x,y)=\eta_{\mu\rho}\eta_{\nu\sigma}G_{(-\hat{\square})}(x,y),
\end{equation}
where we have
\begin{align}
	-\hat{\square}_xG_{-\hat{\square}}(x,y)=\delta^{(d)}(x-y).
\end{align}
Therefore, the first-order deformation simplifies to
\begin{equation}
	S_{[1]}^{(\lambda)}=\frac{\lambda}{2l^2}\int\dif^dx\,\hat{T}_{\mu\nu}\frac{1}{-\hat{\square}}\left(\hat{T}^{\mu\nu}-\frac{1}{d-2}\hat{T}\eta^{\mu\nu}\right)
\end{equation}
on the flat background.\par
By calculating higher-order terms in (\ref{full EOM of h}), in principle, we can determine the deformation at higher orders in the flow parameter $\lambda$. We show the second-order results explicitly in Appendix \ref{appendix second order}.\par
The computation here holds for dimensions $d>2$. It is worth noting that two-dimensional non-local $T\bar{T}$ deformations with an insertion of the operator $F(-\hat{\square})$ have been explored in \cite{Kawamoto:2025oko} to construct traversable AdS wormholes.

\subsection{Effective deformed action}
As we have seen, the non-local deformation of the fixed background induced by Einstein gravity can be derived perturbatively, but its form becomes increasingly complicated at higher orders. Following the spirit of subsection \ref{subsection general proposal}, we provide its corresponding effective deformed action on the deformed metric (i.e., the metric saddle point of $S_{\mathrm{grav}}^{(\lambda)}$).\par
Taking the trace of the Einstein field equation (\ref{Einstein field equation}) and plugging back into the gravitational action (\ref{Einstein gravity}), we can obtain the effective deformed action expressed by the stress tensor of the seed theory (on the metric $g^*_{\mu\nu}$):
\begin{equation}
	S_{\mathrm{EDA}}^{(\lambda)}[g^*,\psi]=\hat{S}[g^*,\psi]+\frac{1}{d-2}\int\dif^dx\sqrt{g^*}\,\hat{T}^*.\label{effective action induced by Einstein gravity}
\end{equation}
To write down the flow equation satisfied by $S_{\mathrm{EDA}}^{(\lambda)}$, we need to express the deformation in terms of the stress tensor of $S_{\mathrm{EDA}}^{(\lambda)}$ itself.\par
From (\ref{effective action induced by Einstein gravity}), the effective stress tensor $T_{\mathrm{EDA}}$ is given by
\begin{equation}
	T_{\mathrm{EDA}}^{(\lambda)\mu\nu}=\hat{T}^{\mu\nu}-\frac{1}{d-2}\hat{T}g^{\mu\nu}-\frac{2}{d-2}\frac{\partial\hat{T}}{\partial g_{\mu\nu}}.\label{effective stress tensor}
\end{equation}
The relation between $T_{\mathrm{EDA}}^{(\lambda)}$ and $\hat{T}$ depends on the specific seed theory due to the presence of the third term. For simplicity, we take the seed theory to be a massless free scalar field,
\begin{align}
	\hat{S}[g,\psi]=\frac{1}{2}\int\dif^dx\sqrt{g}\,\nabla^{\mu}\phi\nabla_{\mu}\phi,
\end{align}
whose stress tensor satisfies $g_{\mu\nu}\frac{\partial\hat{T}}{\partial g_{\mu\nu}}=\hat{T}$. In this case, equation (\ref{effective stress tensor}) implies
\begin{equation}
	\hat{T}=-\frac{d-2}{4}T_{\mathrm{EDA}}^{(\lambda)}.
\end{equation}
Thus, from (\ref{general flow equation}), we obtain the total $\lambda$-derivative of the effective deformed action:
\begin{equation}
	\frac{\dif}{\dif\lambda}S_{\mathrm{EDA}}^{(\lambda)}[g^*,\psi]=\frac{1}{4\lambda}\int\dif^dx\sqrt{g^*}T_{\mathrm{EDA}}^{(\lambda)*},
\end{equation}
where the flow of the saddle-point metric is governed by the Einstein field equation (\ref{Einstein field equation}). It is worth noting that our approach differs from that in \cite{Brizio:2024arr, Morone:2024sdg}. The latter considers coupling a deformed field theory to Einstein gravity, employs the metric approach in \cite{Conti:2022egv} to derive a metric flow from a non-dynamical auxiliary field action, and interprets this metric flow as a Ricci-Bourguignon flow by incorporating the Einstein equation, which holds only when the metric simultaneously satisfies both the metric flow equation and the Einstein equation.

\section{Stress-tensor deformation of scalar field theories on deformed background}\label{sec:scalars}
Starting from this section, we focus on the stress-tensor deformations on the deformed background. Utilizing the Palatini approach in subsection \ref{Subsection 2.2} and \ref{subsection Eigenvalue method}, we investigate the gravitational descriptions of several well-known stress-tensor deformations.

\subsection{Single scalar field theory}
We begin our discussion with the deformation theory of a single scalar field. Assuming that this class of deformation theories involves a unique tensor built from the matter field\footnote{In principle, tensors constructed from a single scalar field may also contain higher-order derivatives, such as $\tilde X_{\mu\nu}=\nabla_{\mu}\nabla_{\nu}\phi$. However, in the present context and subsequent examples, such tensors are excluded from the deformed action.}, $X_{\mu\nu}=\partial_{\mu}\phi\partial_{\nu}\phi$, which is a rank-one symmetric tensor. It is straightforward to verify that products of multiple $X_{\mu\nu}$ tensors can be expressed in terms of $X_{\mu\nu}$ and its trace,
\begin{align}
    (X^n)_{\mu\nu}=(X)^{n-1}X_{\mu\nu}.
\end{align}
The stress tensor can be formally written as $T_{\mu\nu}=a(\phi,X)X_{\mu\nu}+b(\phi,X)g_{\mu\nu}$. One can verify that the products of multiple stress tensors can be expressed as
\begin{align}
    (T^n)_{\mu\nu}=A_n(\phi,\text{tr}T)T_{\mu\nu}+B_n(\phi,\text{tr}T)g_{\mu\nu}.
\end{align}
Hence, in the absence of higher-derivative terms, the multi-trace stress-tensor deformation in a single scalar field theory reduces to a function of the single-trace operator. For certain deformations, the deformed Lagrangian $\mathcal{B}^{(\lambda)}$ takes a closed form \cite{Bonelli:2018kik}, given by a local function of $\lambda$, $\phi$, and $X$. Within the framework of the geometric description, the deformed Lagrangian is derived from a classical $f(\mathcal{R})$ gravitational theory that is non-minimally coupled to the scalar field. In what follows, we analyze specific deformed Lagrangians that admit closed-form expressions, with the aim of constructing the associated gravitational theories.
\subsubsection{Nambu-Goto action in $d$ dimensions}
We begin our discussion with the generalized Nambu-Goto action in $d$ dimensions. In two dimensions, the Nambu-Goto action is obtained via a $T\bar T$ deformation of the free scalar field theory. When extended to higher dimensions, the flow equation of the deformed Lagrangian \cite{Ferko:2024zth, Babaei-Aghbolagh:2024hti} (see also \cite{Brizio:2026ynw}) modifies to 
\begin{align}
    \partial_{\lambda}\mathcal{B}^{(\lambda)}&=\frac{1}{2d}\text{tr}(T^{(\lambda)})^2-\frac{1}{d^2}(\text{tr}T^{(\lambda)})^2\notag\\
    &\quad-\frac{d-2}{2d^{3/2}\sqrt{d-1}}\text{tr}T^{(\lambda)}\sqrt{\text{tr}(T^{(\lambda)})^2-\frac{1}{d}(\text{tr}T^{(\lambda)})^2}.\label{Flow equation of generalized Nambu-Goto action}
\end{align}
When the seed theory is that of a free scalar field, $\mathcal{B}^{(0)}=g^{\mu\nu}\partial_{\mu}\phi\partial_{\nu}\phi=X$, the deformed Lagrangian takes the closed form
\begin{align}\label{free scalar field Nambu-Goto action}
    \mathcal{B}^{(\lambda)}&=\frac{1-\sqrt{1-2\lambda X}}{\lambda}.
\end{align}
As introduced in \cite{Li:2025lpa}, this deformed theory can be reproduced from a classical $f(\mathcal{R})$ gravitational theory with a minimal coupling to the scalar field,
\begin{align}\label{naivemodel}
    \mathcal{A}^{(\lambda)}=(d-1)X+\frac{1}{\lambda(d-1)}\Big(\frac{l^2\mathcal{R}}{2(d-2)^2}-l\sqrt{\mathcal{R}}\Big).
\end{align}
The EoM is derived by varying the above gravitational action with respect to the metric. By choosing the positive branch of the solution, we obtain
\begin{align}
        &\quad\Big(\frac{1}{(d-2)^2}-\frac{1}{l\sqrt{\mathcal{R}}}\Big)l^2\mathcal{R}_{(\mu\nu)}-\Big(\frac{l^2\mathcal{R}}{2(d-2)^2}-l\sqrt{\mathcal{R}}\Big)g_{\mu\nu}\notag\\
    &=-2\lambda (d-1)^2(X_{\mu\nu}-\frac{1}{2}Xg_{\mu\nu}).
\end{align}
Taking the trace of this equation yields
\begin{align}\label{free scalar field sqrt R}
       l\sqrt{\mathcal{R}}&=(d-1)(d-2)(1-\sqrt{1-2\lambda X}).
\end{align}
By plugging (\ref{free scalar field sqrt R}) into the gravitational Lagrangian (\ref{naivemodel}), we can recover the deformed field theory Lagrangian (\ref{free scalar field Nambu-Goto action}). The $\sqrt{\mathcal{R}}$ contribution in (\ref{naivemodel}) signals modified $f(\mathcal{R})$ dynamics; if the scale $l$ is tied to the Hubble radius, the model may naturally account for late-time cosmic acceleration~\cite{Tsujikawa:2007xu, DeFelice:2010aj}, connecting emergent gravity to cosmological implications.\par
The procedure for constructing the above gravitational action admits a natural generalization to a self-interacting scalar field theory. Consider a seed theory whose Lagrangian takes the form $\mathcal{B}^{(0)}=X-2V(\phi)$. The deformed Lagrangian is expressed by the following closed form
\begin{align}\label{self interecting scalar field Nambu-Goto action}
    \mathcal{B}^{(\lambda)}=\frac{1-2\lambda V-\sqrt{1-2\lambda (1-\lambda V)X}}{\lambda (1-\lambda V)}.
\end{align}
From the perspective of geometric description, the following $f(\mathcal{R})$ gravitational theory with a non-minimal coupling to the scalar field is postulated,
\begin{align}
    \mathcal{A}^{(\lambda)}&=\frac{2 (d-1)(1-\lambda  V)X}{k(V)}+\frac{k(V)}{4\lambda(d-1)}\Big(\frac{l^2\mathcal{R}}{(d-2)^2}-\frac{2l\sqrt{\mathcal{R}}}{1-\lambda  V}\Big)\notag\\
    &\quad+\frac{d-2}{4\lambda(1-\lambda V)}\Big(\frac{k(V)}{1-\lambda V}-2(1-2\lambda V)\Big),\label{gravitational action of the generalized NG action}
\end{align}
where $k(V)=d(1-\lambda V)(1-2\lambda V)-\sqrt{(1-\lambda V)^2(4-4d+d^2(1-2\lambda V)^2)}$. Taking the variation with respect to the metric, we find
\begin{align}\label{relation of SqR}
    \sqrt{\mathcal{R}}&=\frac{(d-1)(d-2)(k(V)-2(1-\lambda V)\sqrt{1-2\lambda (1-\lambda V)X}}{l(1-\lambda V)k(V)}.
\end{align}
Then, we can substitute it into (\ref{gravitational action of the generalized NG action}) to reproduce the deformed field theory Lagrangian (\ref{self interecting scalar field Nambu-Goto action}). When $V(\phi)$ is constant, the last term of equation~(\ref{gravitational action of the generalized NG action}) shows that the deformation generates a volume term, yielding a cosmological constant as an intrinsic vacuum contribution. Thus, the cosmological constant arises dynamically rather than by input, offering a natural and calculable origin with direct implications for dark energy and cosmic acceleration.\par
At the end of this subsection, we derive the dynamical equation governing the deformed metric $g^{(\lambda)}$. Our discussion is based on the Palatini formalism, where the action depends on both the metric and an independent connection. Variation with respect to the connection yields an additional EOM. By combining it with the EOM obtained from metric variation and eliminating the connection, we find that the deformed metric satisfies eq. (\ref{modified Einstein's equation for single scalar field}). By taking derivatives of (\ref{self interecting scalar field Nambu-Goto action}), and using the relation (\ref{relation of SqR}), we obtain
\begin{align}
    \mathcal{A}^{(\lambda)}|_{g=g^{(\lambda)}}&=\frac{1-2\lambda V-\sqrt{1-2\lambda (1-\lambda V)X}}{\lambda (1-\lambda V)},\ \ \ \mathcal{A}^{(\lambda)}_X|_{g=g^{(\lambda)}}=\frac{2 (d-1)(1-\lambda  V)}{k(V)},\notag\\
    \mathcal{A}^{(\lambda)}_{\mathcal{R}}|_{g=g^{(\lambda)}}&=\frac{l^{2}k(V)(k(V) - 2(d-1)(1-\lambda V)\sqrt{1-2\lambda(1-\lambda V)X})}{4\lambda(d-2)^{2}(d-1)^{2}(k(V) - 2(1-\lambda V)\sqrt{1-2\lambda(1-\lambda V)X})}.
\end{align}
Substituting these expressions into eq. (\ref{modified Einstein's equation for single scalar field}) yields a differential equation that involves only the matter field and the deformed metric.

\subsubsection{$\det T$ deformation in $d$ dimensions}
Another interesting class of stress-tensor deformation in higher dimensions is the $(-\det T)^{1/\alpha}$ deformation introduced in Refs. \cite{Cardy:2018sdv,Bonelli:2018kik}. The flow equation of the field theory Lagrangian is given by
\begin{align}\label{flow equation of Lagrangian in detT deformation}
    \partial_{\lambda}\mathcal{B}^{(\lambda)}&=\frac{1}{\alpha-d}\Big(-\det T^{(\lambda)}\Big)^{\frac{1}{\alpha}},
\end{align}
where $\alpha$ is a real parameter. When the seed theory is a single free scalar field theory, $\mathcal{B}^{(0)}=(1/2)g^{\mu\nu}\partial_{\mu}\phi\partial_{\nu}\phi=X/2$, the deformed Lagrangian admits a closed form for certain values of $\alpha$. By introducing the redefinitions $\mathcal{B}^{(\lambda)}=\sqrt{X}f^{\frac{1}{d-\alpha}}$ and $Y=(-1)^{\alpha}X^{\frac{\alpha-d}{2}}$, the deformed Lagrangian satisfies the following PDE \cite{Bonelli:2018kik},
\begin{align}\label{PDE in detT deformation}
    (\partial_{\lambda}f)^\alpha+f^{\alpha}\partial_{Y}f=0.
\end{align}\par
When $\alpha=1$, (\ref{flow equation of Lagrangian in detT deformation}) corresponds to the $\det T$ deformation, and eq.(\ref{PDE in detT deformation}) reduces to the Burger's equation. The deformed Lagrangian takes a closed form,
\begin{align}\label{closed form of Lagrangian of detT deformation}
\mathcal{B}^{(\lambda)}|_{\alpha=1}=\Big(\frac{\sqrt{1+4\lambda(X/2)^{d-1}}-1}{2\lambda}\Big)^{\frac{1}{d-1}}.
\end{align}
Now, we construct the gravitational theory corresponding to the above deformed Lagrangian. Note that the structure inside the parentheses in (\ref{closed form of Lagrangian of detT deformation}) resembles that appearing in the Nambu-Goto action. We can consider an $f(\mathcal{R})$ gravity minimally coupled to a free scalar field to reproduce it. The gravitational Lagrangian is specified as 
\begin{align}
    \mathcal{A}^{(\lambda)}=\Big[\Big(\frac{X}{2}\Big)^{d-1}+\frac{1}{\lambda}\Big(-\frac{(d-2)^2}{4}+\frac{1}{d-1}\mathcal{R}^{\frac{d-1}{2}}-\frac{1}{(d-1)^2(d-2)^2}\mathcal{R}^{d-1}\Big)\Big]^{\frac{1}{d-1}}.\label{TTbarlike for single scalar gravitational action}
\end{align}
Taking variation with respect to the metric, we find
\begin{align}
    \mathcal{R}^{\frac{d-1}{2}}&=\frac{(d-1)(d-2)}{2}\Big(d-1-\sqrt{1+4\lambda(X/2)^{d-1}}\Big).\label{TTbarlike for single scalar EoM for Ricci scalar}
\end{align}
By plugging this expression into (\ref{TTbarlike for single scalar gravitational action}), we can recover the deformed field theory Lagrangian (\ref{closed form of Lagrangian of detT deformation}).\par
As another example, for $\alpha=2$ — corresponding to the $d$-dimensional $\sqrt{-\det T}$ deformation — the deformed Lagrangian in a closed form is given by \cite{Bonelli:2018kik}
\begin{align}
    \mathcal{B}^{(\lambda)}|_{\alpha=2}=\Big[\Big(\frac{X}{2}\Big)^{2-d}+\lambda\Big(\frac{X}{2}\Big)^{\frac{2-d}{2}}\Big]^{\frac{1}{2-d}}.
\end{align}
The corresponding gravitational Lagrangian can be constructed as
\begin{align}
    \mathcal{A}^{(\lambda)}=\Big[\frac{d}{2(d-1)}\Big(\frac{X}{2}\Big)^{2-d}+\frac{|b(\lambda)|}{\lambda}\Big(\mathcal{R}^{\frac{d(2-d)}{2}}+\frac{|b(\lambda)|(d-2)}{2\lambda^3(d-1)}\mathcal{R}^{d(2-d)}\Big)\Big]^{\frac{1}{2-d}},
\end{align}
where $b(\lambda)$ is an arbitrary function with the correct physical dimension. By using the EOM of the Palatini Ricci scalar, we obtain
\begin{align}
    \mathcal{R}^{d(2-d)}=\frac{\lambda^4}{b(\lambda)^2}\Big(\frac{X}{2}\Big)^{2-d}.
\end{align}
Then, we can plug it into $\mathcal{A}^{(\lambda)}$ to recover the deformed field theory Lagrangian.\par
Using the method outlined in subsection \ref{Subsection 2.2}, we can derive the EOM governing the deformed metric. Taking the $\det T$ deformation as an example, we combine (\ref{TTbarlike for single scalar gravitational action}) and (\ref{TTbarlike for single scalar EoM for Ricci scalar}) to obtain
\begin{align}
    \mathcal{A}^{(\lambda)}_{\mathcal{R}}|_{g=g^{(\lambda)}}&=\frac{[ (d-2)(d-1)(d-1 - \sqrt{1 + 2^{3 - d} X^{d - 1}\lambda}) ]^{\frac{d - 3}{d - 1}}(-1 + \sqrt{1 + 2^{3 - d} X^{d - 1}\lambda})^{\frac{1}{d - 1}}}{2^{\frac{d-2}{d - 1}}\lambda^{\frac{1}{d-1}}(d - 2)(d - 1)},\notag\\
    \mathcal{A}^{(\lambda)}_{X}|_{g=g^{(\lambda)}}&=2^{- \frac{d^2-3d+3}{d - 1}}X^{d - 2}\Big(\frac{-1 + \sqrt{1 + 2^{3 - d} X^{d - 1}\lambda}}{\lambda}\Big)^{-\frac{d - 2}{d - 1}}.
\end{align}
One can further plug this expression into (\ref{modified Einstein's equation for single scalar field}) to obtain the differential equation for the deformed metric $g^{(\lambda)}$.

\subsection{Multiple scalar field theory}
Next, we consider theories involving multiple scalar fields. The $T\bar{T}$ and root-$T\bar{T}$ deformation of $N$ free scalars in two dimensions have been well studied \cite{Smirnov:2016lqw,Babaei-Aghbolagh:2022leo,Babaei-Aghbolagh:2025lko}. In this subsection, we investigate their corresponding gravitational descriptions in our framework. Additionally, we consider the (root-)$T\bar{T}$-like deformation introduced in \cite{Babaei-Aghbolagh:2024hti,Li:2025lpa} as another example.
\subsubsection{Scalar ModMax theory in two dimensions}\label{subsection SMM}
Firstly, we consider the theory of $N$ free scalars,
\begin{equation}
	\mathcal{L}=\frac{1}{2}\sum_{i,j}G_{ij}\partial_\mu\phi^i\partial^\mu\phi^j\equiv\frac{X}{2},
\end{equation}
where $i,j=1,2,\cdots,N$ and $G_{ij}$ is the metric on the target space. The root-$T\bar{T}$ deformed version of this theory is known as the Scalar ModMax (SMM) theory\cite{Babaei-Aghbolagh:2022leo, Babaei-Aghbolagh:2025lko}:
\begin{equation}
	\mathcal{B}^{(\gamma)}_{\mathrm{SMM}}=\frac{1}{2}\left(\cosh(\gamma)X-\sinh(\gamma)\sqrt{2X_2-X^2}\right).\label{scalar ModMax}
\end{equation}
To construct its corresponding gravitational Lagrangian, we again employ the Palatini $f(\mathcal{R})$ formalism. The total Lagrangian is assumed to consist of three parts: a matter sector proportional to the seed theory, an $f(\mathcal{R})$ gravity sector, and a non-minimal coupling between gravity and the matter fields. Formally, it can be written as 
\begin{align}
    \mathcal{A}^{(\gamma)}=a(\gamma)X+f(\mathcal{R},\gamma)+u(\mathcal{R},X,X_2,\gamma).\label{Non-minimal gravity for deformed N scalars}
\end{align}
Taking the variation of the action with respect to the metric, we have
\begin{align}
	2(\partial_{\mathcal{R}}f+\partial_{\mathcal{R}}u)\mathcal{R}_{(\mu\nu)}+(2a+2\partial_{X}u)X_{\mu\nu}+4\partial_{X_2}u(X^2)_{\mu\nu}-\mathcal{A}^{(\gamma)}g_{\mu\nu}=0.
\end{align}
The trace of this equation is
\begin{align}
	(\partial_{\mathcal{R}}f+\partial_{\mathcal{R}}u)\mathcal{R}+X\partial_{X}u+2X_2\partial_{X_2}u-f-u=0.\label{Ricci scalar EOM in scalar theory}
\end{align}
Given that the Scalar ModMax Lagrangian (\ref{scalar ModMax}) is a homogeneous function of the metric, it is natural to incorporate a structurally analogous term into the gravitational action, as this homogeneity is preserved under metric variation. As an illustrated example, the explicit form of the gravitational Lagrangian is given by
\begin{equation}
	\mathcal{A}^{(\gamma)}=\frac{1}{2}\cosh\gamma X+b(\gamma)\mathcal{R}^p+c(\gamma)\mathcal{R}^q(2X_2-X^2)^s,\label{ansatz for gravitational action of scalar modmax}
\end{equation}
Eq. (\ref{Ricci scalar EOM in scalar theory}) then gives
\begin{equation}
	\mathcal{R}^{p-q}=\frac{2s+q-1}{1-p}\frac{c(\gamma)}{b(\gamma)}(2X_2-X^2)^s.
\end{equation}
Plugging it back into (\ref{ansatz for gravitational action of scalar modmax}), to reproduce the Scalar ModMax theory, there should be
\begin{equation}
	\begin{split}
		q&=p(1-2s),\\
		c(\gamma)&=(-1)^{2s}(4s)^{-2s}(2s-1)^{2s-1}b^{1-2s}(\sinh\gamma)^{2s}.
	\end{split}
\end{equation}
To avoid the multivaluedness of the solution of $\mathcal{R}$, we can set $s=\frac{1}{2p}$. Then we have
\begin{equation}
	\begin{split}
		q&=p-1,\\
		c(\gamma)&=\left(-\frac{1}{2}\right)^{\frac{1}{p}}p\left(b^{p-1}(1-p)^{1-p}\sinh\gamma\right)^{\frac{1}{p}}.
	\end{split}\label{c in SMM}
\end{equation}
One can always choose $c(\gamma)$ to be real. Thus, the gravitational action is constructed as
\begin{equation}
	\mathcal{A}^{(\gamma)}=\frac{1}{2}\cosh\gamma X+b(\gamma)\mathcal{R}^p+c(\gamma)\mathcal{R}^{p-1}(2X_2-X^2)^\frac{1}{2p}.\label{gravitational action of scalar modmax}
\end{equation}
where $p\neq1$, $b(\gamma)$ is an arbitrary function factor independent of $g_{\mu\nu}$ and $\phi$ with the correct physical dimension and $c(\gamma)$ is given by (\ref{c in SMM}). Substituting the equation
\begin{equation}
	\mathcal{R}=\frac{1-p}{p}\frac{c(\gamma)}{b(\gamma)}(2X_2-X^2)^{\frac{1}{2p}}
\end{equation}
back into (\ref{gravitational action of scalar modmax}), we can reproduce the Scalar ModMax theory on the deformed background $g^{(\lambda)}$. The flow equation satisfied by $g^{(\lambda)}$ can be obtained from the modified Einstein's equation (\ref{modified Einstein's equation in 2d}).

An alternative systematic approach for constructing the gravitational action is the eigenvalue method introduced in subsection \ref{subsection Eigenvalue method}. By employing the diagonalization (\ref{diagonalization}), the SMM Lagrangian can be expressed in terms of eigenvalues,
\begin{align}\label{scalar modmax}
	\mathcal{B}_{\mathrm{SMM}}^{(\gamma)}&=\frac{1}{2}\left(\cosh(\gamma)(\chi_1+\chi_2)-\sinh(\gamma)|\chi_1-\chi_2|\right)\nonumber\\
	&=\frac{1}{2}\left(e^{-\gamma}\chi_1+e^\gamma\chi_2\right).
\end{align}
Assuming that the corresponding gravitational Lagrangian takes a simple form,
\begin{equation}\label{gravitational Lagrangian for SMM in eigenvalues}
	\mathcal{A}^{(\gamma)}=a(\gamma)(\chi_1+\chi_2)+\sum_{j=1}^2b_j(\gamma)r_j^{q_j(\gamma)},
\end{equation}
the metric EOM (\ref{Eigenvalue method EoM for rj})	then gives
\begin{align}
	b_1r_1^{q_1}=\frac{a(1-q_2^{-1})(\chi_2-\chi_1)}{q_1(2-\sum_{j=1}^2q_j^{-1})},\quad b_2r_2^{q_2}=\frac{a(1-q_1^{-1})(\chi_1-\chi_2)}{q_2(2-\sum_{j=1}^2q_j^{-1})}.\label{ricci tensor EOM in SMM}
\end{align}
By substituting the algebraic equations (\ref{ricci tensor EOM in SMM}) back into (\ref{gravitational Lagrangian for SMM in eigenvalues}), one can eliminate the Ricci tensor eigenvalues $r_1,r_2$ in $\mathcal{A}^{(\gamma)}$ to recover the deformed Lagrangian $\mathcal{B}^{(\gamma)}$. For this result to match the Scalar ModMax theory (\ref{scalar modmax}), the parameters $a(\gamma)$ and $q_j(\gamma),j=1,2$ must satisfy the constraints
\begin{equation}
	\frac{1-q_2^{-1}}{1-q_1^{-1}}=e^{2\gamma},\quad a=\frac{1}{2}\cosh(\gamma).
\end{equation}

\subsubsection{Scalar Born-Infeld theory in two dimensions}\label{subsection BI2d}
The $T\bar{T}$-deformed theory of $N$ free scalars is known as the two-dimensional Born-Infeld theory \cite{Cavaglia:2016oda}
\begin{equation}
	\mathcal{B}^{(\lambda)}_{\mathrm{BI-2d}}=\frac{2}{\lambda}\left(1-\sqrt{1-\frac{1}{2}\lambda X+\frac{1}{8}\lambda^2(X^2-X_2)}\right).\label{2d BI theory}
\end{equation}
We still assume that the corresponding gravitational Lagrangian can be constructed in the form of (\ref{Non-minimal gravity for deformed N scalars}). In contrast to the SMM case, the Born-Infeld Lagrangian is not a homogeneous function of the metric. As a result, constructing a closed gravitational action that could reproduce (\ref{2d BI theory}) becomes highly non-trivial. A useful technique to address this issue is to introduce an additional parameter, $\epsilon$. When evaluated at the metric saddle point, the gravitational action yields the Born-Infeld theory at the leading order in $\epsilon$. The gravitational Lagrangian can be specifically constructed as
\begin{align}\label{gravitational action of 2d BI theory}
	\mathcal{A}^{(\lambda)}&=\frac{X}{2}+|b(\lambda)|l^{2\epsilon}\mathcal{R}^{1+\epsilon}\left(\epsilon^{-\frac{1}{2}}(1+\epsilon)-\frac{\lambda}{4}|b(\lambda)|\mathcal{R}\right)\nonumber\\
	&\quad+l^{2\epsilon}\frac{X}{2}\mathcal{R}^\epsilon\left((\sqrt{2}-1)(1+\epsilon)-\frac{\sqrt{2}}{2}\lambda|b(\lambda)|\epsilon^{\frac{1}{2}}\mathcal{R}\right)-\frac{\lambda}{8}\epsilon l^{2\epsilon} X_2\mathcal{R}^\epsilon,
\end{align}
where $b(\lambda)$ is an arbitrary function with the correct physical dimension. The EOM (\ref{Ricci scalar EOM in scalar theory}) then gives
\begin{equation}
	\mathcal{R}=\frac{4-\sqrt{2}\lambda X-2\sqrt{1-\frac{1}{2}\lambda X+\frac{1}{8}\lambda^2(X^2-X_2)}}{\lambda\epsilon^{-\frac{1}{2}}|b(\lambda)|}.
\end{equation}
Plugging it back into (\ref{gravitational action of 2d BI theory}) and expanding the result in powers of $\epsilon$, we can reproduce the two-dimensional Born-Infeld Lagrangian
\begin{equation}
	\mathcal{A}^{(\lambda)}|_{g=g^{(\lambda)}}=\frac{2}{\lambda}\left(1-\sqrt{1-\frac{1}{2}\lambda X+\frac{1}{8}\lambda^2(X^2-X_2)}\right)+O(\epsilon).
\end{equation}

In contrast, the eigenvalue method does not require introducing an additional parameter. By employing the diagonalization (\ref{diagonalization}), the two-dimensional Born-Infeld Lagrangian simplifies to
\begin{align}
	\mathcal{L}_{\mathrm{BI-2D}}&=\frac{2}{\lambda}\left(1-\sqrt{1-\frac{1}{2}\lambda\sum_{j=1}^2\chi_j+\frac{1}{8}\lambda^2\left(\Big(\sum_{j=1}^2\chi_j\Big)^2-\sum_{j=1}^2\chi_j^2\right)}\right)\nonumber\\
	&=\frac{2}{\lambda}-\frac{2}{\lambda}\prod_{j=1}^2\left(1-\frac{\lambda}{2}\chi_j\right)^{\tfrac{1}{2}}.\label{2d BI theory in terms of eigenvalues}
\end{align}
Following the method in subsection \ref{subsection Eigenvalue method}, one can construct the corresponding gravitational Lagrangian as
\begin{align}
	\mathcal{A}^{(\lambda)}&=\mathcal{A}_0^{(\lambda)}-\frac{2}{\lambda}\prod_{j=1}^2\left(\left(a_j\chi_j^{\frac{1}{2}}+b_jr_j^{\frac{1}{2}}-1\right)-\frac{\lambda}{2}\chi_j\right)^{\frac{1}{2}},\label{gravitational Lagrangian of 2d BI theory}
\end{align}
where
\begin{equation}
	\mathcal{A}_0^{(\lambda)}=\frac{1}{2\lambda}\prod_{j=1}^2\left(a_j\chi_j^{\frac{1}{2}}+b_jr_j^{\frac{1}{2}}\right)
\end{equation}
satisfies
\begin{equation}
	2r_j\partial_{r_j}\mathcal{A}_0^{(\lambda)}+2\chi_j\partial_{\chi_j}\mathcal{A}_0^{(\lambda)}-\mathcal{A}_0^{(\lambda)}=0,\quad j=1,2,
\end{equation}
with parameters $a_j,b_j$ independent of $\chi_j$ and $r_j$. The metric EOM (\ref{Eigenvalue method EoM for rj}) then gives
\begin{equation}
	r_j^{\frac{1}{2}}=-\frac{a_j}{b_j}\chi_j^{\frac{1}{2}}+\frac{2}{b_j},\quad j=1,2.\label{EOM 1 in terms of eigenvalues in 2d BI theory}
\end{equation}
Substituting (\ref{EOM 1 in terms of eigenvalues in 2d BI theory}) into (\ref{gravitational Lagrangian of 2d BI theory}), one can recover the two-dimensional Born-Infeld Lagrangian (\ref{2d BI theory in terms of eigenvalues}) correctly. In particular, the constant term $2/\lambda$ in (\ref{2d BI theory in terms of eigenvalues}) emerges directly from the $\mathcal{A}_0^{(\lambda)}$ part of the construction.

\subsubsection{$T\bar T$-like deformation of Nambu-Goto string in $d$ dimensions}
To conclude this section, we examine a special class of $T\bar T$-like deformed theories in $d$ dimensions introduced in \cite{Babaei-Aghbolagh:2024hti, Li:2025lpa}. Suppose the field theory contains only one independent tensor constructed from the matter fields, denoted as $X_{\mu\nu}=G_{ij}(\psi)\partial_{\mu}\phi^{(i)}\partial_{\nu}\phi^{(j)}$, where $G_{ij}$ is the metric in the target space. The deformed field theory Lagrangian is written in terms of the eigenvalues of $X_{\mu\nu}$,
\begin{align}
    \mathcal{B}^{(\lambda)}(\chi_j,\psi)=\mathcal{B}_0(\chi_j,\psi)+\lambda^{1-\Sigma}l^{\Delta}\prod_{j=1}^{d}(\chi_j^{p_j/2}-\beta_j^{p_j/2})^{1/p_j},\label{non-minimal effective deformed action}
\end{align}
where $\lbrace{\beta_j}\rbrace$ is the deformation parameters of the root-$T\bar T$-like operator as discussed in \cite{Babaei-Aghbolagh:2024hti}, $\lbrace{p_j}\rbrace$ are the numbers that characterizing the deformation, $\Sigma=\sum_{j=1}^{d}1/p_j$, and $\Delta=(2\Sigma+d-4)d/2$. The redundant part $\mathcal{B}_0$, which may also be interpreted as the seed theory Lagrangian when $\Sigma<1$, satisfies the differential equations $2\chi_j\partial_{\chi_j}\mathcal{B}_0-\mathcal{B}_0=0$ for $j=1,...,d$. The solution is given by
\begin{align}
\mathcal{B}_0=C(\psi)\prod_{j=1}^{d}\chi_j^{1/2}=C(\psi)\frac{\sqrt{\det(X_{\mu\nu})}}{\sqrt{\det(g_{\mu\nu})}}, \label{NG action}
\end{align}
 which is analogous to the Lagrangian of the Nambu-Goto string \cite{Polchinski:1998rq} up to a regular function $C(\psi)$. The stress tensor of the deformed field theory is entirely determined by the second term in (\ref{non-minimal effective deformed action}), and its eigenvalues $\lbrace{\tau_j}\rbrace$ are formulated as 
 \begin{align}
\tau_j^{(\lambda)}&=2\chi_j\partial_{\chi_j}\mathcal{B}^{(\lambda)}-\mathcal{B}^{(\lambda)}\notag\\
&=\lambda^{1-\Sigma}l^{\Delta}\frac{\beta_j^{p_j/2}}{\chi_j^{p_j/2}-\beta_j^{p_j/2}}\prod_{k=1}^{d}(\chi_k^{p_k/2}-\beta_k^{p_k/2})^{1/p_k}.\label{non-minimal effective stress tensor 3}
 \end{align}
By plugging it into (\ref{non-minimal effective deformed action}), we obtain the flow equation of the deformed field theory action,
\begin{align}
        \partial_{\lambda}S_{\text{EDA}}^{(\lambda)}&=\int d^dx\sqrt{g}\partial_{\lambda}\mathcal{B}^{(\lambda)}\notag\\
        &=(1-\Sigma)l^{\frac{\Delta}{1-\Sigma}}b^{\frac{1}{2(1-\Sigma)}}\int d^dx\sqrt{g}\Big(\prod_{j=1}^{d}(\tau_j^{(\lambda)})^{1/p_j}\Big)^{\frac{1}{\Sigma-1}},
\end{align}
where $b=\prod_{j=1}^{d}\beta_j$. In particular, if we set $p_1=p_2=...=p_d=p$, the deformation operator reduces to the determinant of the stress tensor, $\mathcal{O}_{\text{st}}=(\det[(T^{(\lambda)}_{\text{EDA}})^{\mu}_{\nu}])^{\frac{1}{d-p}}$ \cite{Bonelli:2018kik, Cardy:2018sdv}.\par
Next, we construct the gravitational Lagrangian capable of reproducing the deformed field theory Lagrangian given in (\ref{non-minimal effective deformed action}), which is explicitly formulated as
\begin{align}
    \mathcal{A}^{(\lambda)}&=\mathcal{B}_0+\lambda^{1-\Sigma}l^{\Delta}\prod_{j=1}^{d}\Big(\chi_j^{p_j/2}-(\mathfrak{p}_{j}\chi_{j}^{\mathfrak{q}_j}+\mathfrak{s}_jr_{j}^{\mathfrak{q}_j}+\frac{\mathfrak{q}_j-1}{\mathfrak{q}_j}\beta_j)^{p_j/2}\Big)^{1/p_j},\label{non-minimal gravitational action}
\end{align}
where $\mathfrak{p}_{j}$, $\mathfrak{q}_{j}$, and $\mathfrak{s}_{j}$ are some parameters independent of $\chi$ and $r$. By employing the metric EOM calculated in (\ref{Eigenvalue method EoM for rj}), we have
\begin{align}
    r_j^{\mathfrak{q}_j}=-\frac{\mathfrak{p}_j}{\mathfrak{s}_j}\chi_j^{\mathfrak{q}_j}+\frac{\beta_j}{\mathfrak{q}_j\mathfrak{s_j}},\ \ \ \ \ (\mathfrak{q}_j\neq 1).\label{non-minimal EoM R}
\end{align}
One can further plug this equation into (\ref{non-minimal gravitational action}) to recover the deformed field theory Lagrangian.\par
In the following, we use the above formulas to analyze a specific example in two dimensions. The gravitational action is constructed as follows,
\begin{align}
    S^{(\lambda)}_{\text{grav}}&=\int d^2x\sqrt{g}\Bigg\lbrace\mathcal{B}_0+\frac{1}{\lambda l^{2}}\prod_{\sigma=\pm}\bigg(\sqrt{\frac{1}{2}\Big(X+\sigma\sqrt{2X^{\mu}_{\nu}X^{\nu}_{\mu}-X^2}}\Big)\notag\\
    &\quad-\sqrt{M^2+l^{-1}\sqrt{\mathcal{R}+\sigma\sqrt{2\mathcal{R}^{\mu}_{\nu}\mathcal{R}^{\nu}_{\mu}-\mathcal{R}^2}}+l^{-1}\sqrt{X+\sigma\sqrt{2X^{\mu}_{\nu}X^{\nu}_{\mu}-X^2}}}\bigg)\Bigg\rbrace. \label{D=2 gravitational action}
\end{align}
By setting $p_1=p_2=1$, $\mathfrak{p}_1=\mathfrak{p}_2=\mathfrak{s}_1=\mathfrak{s}_2=1$, $\mathfrak{q}_1=\mathfrak{q}_2=\frac{1}{2}$, and $\beta_1=\beta_2=-\frac{\sqrt{2}M^2}{2}$, the action (\ref{non-minimal gravitational action}) can be identified with (\ref{D=2 gravitational action}). From the metric EOM, we obtain the explicit form of the Ricci curvature tensor,
\begin{align}
    \mathcal{R}^{\mu}_{\nu}=(1+\frac{2\sqrt{2}M^2l}{\sqrt{\chi_1}+\sqrt{\chi_2}})X^{\mu}_{\nu}+\sqrt{2}M^2l(\sqrt{2}M^2l+\sqrt{\chi_1}+\sqrt{\chi_2}-\frac{X}{\sqrt{\chi_1}+\sqrt{\chi_2}})\delta^{\mu}_{\nu},
\end{align}
where $\chi_1=\frac{1}{2}(X+\sqrt{2X^{\mu}_{\nu}X^{\nu}_{\mu}-X^2})$, and $\chi_2=\frac{1}{2}(X-\sqrt{2X^{\mu}_{\nu}X^{\nu}_{\mu}-X^2})$. The reproduced deformed field theory satisfies the flow equation
\begin{align}
    \partial_{\lambda}S^{(\lambda)}_{\text{EDA}}=\frac{\sqrt{2}}{2M^2l^2}\int d^2x\sqrt{g}\Big[(T_{\text{EDA}})^{\mu}_{\nu}(T_{\text{EDA}})^{\nu}_{\mu}-(T_{\text{EDA}})^2\Big],
\end{align}
which corresponds to that of the $T\bar T$ deformation \cite{Zamolodchikov:2004ce,Smirnov:2016lqw,Cavaglia:2016oda,He:2025ppz}.

\section{Deformation of 4d Maxwell's theory}~\label{sec:gauge}
In this section, we turn to study the $T\bar{T}$-type and root-$T\bar{T}$-type deformations of Maxwell's theory, which, as is well known, correspond to Born-Infeld theory \cite{born1934foundations, Conti:2018jho} and ModMax theory \cite{Bandos:2020jsw, Babaei-Aghbolagh:2022uij}. These models have played important roles in the study of non-linear electrodynamics. The Scalar Born-Infeld theory and the Scalar ModMax theory considered in subsection \ref{subsection SMM} and \ref{subsection BI2d} can be obtained via a dimensional reduction from the 4d vector theories \cite{barbashov1966scattering, Conti:2022egv, Babaei-Aghbolagh:2025uoz}, leading to closely related forms for their corresponding gravitational theories.

\subsection{Root $T\bar T$ deformation}
The root-$T\bar{T}$ deformed theory of 4d Maxwell's theory is the ModMax theory \cite{Babaei-Aghbolagh:2022uij, Borsato:2022tmu, Babaei-Aghbolagh:2022itg, Ferko:2023ozb}. The Lagrangian of $4d$ ModMax theory takes the form
\begin{align}\label{4d ModMax Lagrangian}
    \mathcal{B}^{(\gamma)}_{\text{MM}}=\cosh{\gamma}\ \mathcal{S}+\sinh\gamma \sqrt{\mathcal{S}^2+\mathcal{P}^2},
\end{align}
where $\mathcal{S}=-\frac{1}{4}F_{\mu\nu}F^{\mu\nu}$ and $\mathcal{P}=-\frac{1}{4}F_{\mu\nu}\tilde F^{\mu\nu}$ with $\tilde F^{\mu\nu}=\frac{1}{2}\epsilon^{\mu\nu\rho\sigma}F_{\rho\sigma}$. Here $\epsilon$ is the Levi-Civita tensor that satisfies $\epsilon_{\mu\nu\rho\sigma}=\sqrt{g}\varepsilon_{\mu\nu\rho\sigma}$, where $\varepsilon_{\mu\nu\rho\sigma}$ is the Levi-Civita symbol. Consequently, by using the identity $\varepsilon^{\mu\nu\rho\sigma}\varepsilon_{\alpha\beta\gamma\delta}=\delta^{\mu\nu\rho\sigma}_{\alpha\beta\gamma\delta}$, we have $\epsilon^{\mu\nu\rho\sigma}\epsilon_{\alpha\beta\gamma\delta}=-\delta^{\mu\nu\rho\sigma}_{\alpha\beta\gamma\delta}$ in Lorentzian signature. The above Lagrangian can thus be written as
\begin{align}
	\mathcal{B}^{(\gamma)}_{\text{MM}}=\frac{\cosh\gamma}{4}F_{\mu\nu}F^{\nu\mu}+\frac{\sinh\gamma}{4}\sqrt{4F_{\mu\nu}F^{\nu\rho}F_{\rho\sigma}F^{\sigma\mu}-(F_{\mu\nu}F^{\nu\mu})^2}.
\end{align}

\subsubsection{Non-minimal $f(\mathcal{R})$ gravity}
The simplest approach to constructing the gravitational Lagrangian is to use the Ricci scalar in the Palatini formalism. Similar to the scalar case, the Lagrangian is taken to be decomposed into three parts: a matter sector proportional to the four-dimensional Maxwell theory, an $f(\mathcal{R})$ gravity sector, and a non-minimal coupling between gravity and the matter fields. The Lagrangian can be formally written as 
\begin{align}\label{Non-minimal gravity for deformed Maxwell}
\mathcal{A}^{(\gamma)}=a(\gamma)\mathcal{S}+f(\mathcal{R},\gamma)+u(\mathcal{R},F_2,F_4,\gamma),
\end{align}
where $F_2=F_{\mu\nu}F^{\nu\mu}$, and $F_4=F_{\mu\nu}F^{\nu\rho}F_{\rho\sigma}F^{\sigma\mu}$. Taking the variation of the action with respect to the metric, we have
\begin{align}
2(\partial_{\mathcal{R}}f+\partial_{\mathcal{R}}u)\mathcal{R}_{(\mu\nu)}+(a+4\partial_{F_2}u)(F^2)_{\mu\nu}+8\partial_{F_4}u(F^4)_{\mu\nu}-\mathcal{A}^{(\gamma)}g_{\mu\nu}=0.
\end{align}
The trace of this equation is
\begin{align}\label{metric EoM for ModMax}
(\partial_{\mathcal{R}}f+\partial_{\mathcal{R}}u)\mathcal{R}+2F_2\partial_{F_2}u+4F_4\partial_{F_4}u-2f-2u=0.
\end{align}
In principle, the expression of $\mathcal{R}$ in terms of $(F_2, F_4)$ is obtained from the above trace equation, which can then be substituted into $\mathcal{A}^{(\gamma)}$ to find the effective Lagrangian.\par
The explicit construction here is parallel to the scalar case (\ref{ansatz for gravitational action of scalar modmax}),
\begin{align}
    \mathcal{A}^{(\gamma)}=\cosh\gamma\mathcal{S}+b(\gamma)\mathcal{R}+c(\gamma)\mathcal{R}^{1-2q}(\mathcal{S}^2+\mathcal{P}^2)^q,
\end{align}
where $c(\gamma)=(2q)^{-2q}(2q-1)^{2q-1}b^{1-2q}(\sinh\gamma)^{2q}$, and $b(\gamma)$ is an arbitrary function with the correct physical dimension. To avoid the multivaluedness of the solution of $\mathcal{R}$, we can set $q=-\frac{1}{2}$, which subsequently yields
\begin{align}
    \mathcal{A}^{(\gamma)}&=\cosh\gamma\mathcal{S}+b(\gamma)\mathcal{R}-\frac{b(\gamma)^2}{4\sinh\gamma}\frac{\mathcal{R}^{2}}{\sqrt{\mathcal{S}^2+\mathcal{P}^2}}.
\end{align}
Taking the variation with respect to the metric, we find
    \begin{align}
    \mathcal{R}&=\frac{2\sinh\gamma}{b(\gamma)}\sqrt{\mathcal{S}^2+\mathcal{P}^2}.
\end{align}
By plugging it into the gravitational Lagrangian, we can recover the Lagrangian (\ref{4d ModMax Lagrangian}) of the ModMax theory in four dimensions.
\subsubsection{Eigenvalue method}
Similarly, we also consider constructing the gravitational action using the eigenvalue method. For simplicity, we introduce a tetrad field $e_{\mu}^{a}$ defined by the relation $g_{\mu\nu}=e_{\mu}^{a}e_{\nu}^{b}\eta_{ab}$. The electromagnetic field tensor expressed in the tetrad basis is given by $F_{ab}=e^{\mu}_{a}e^{\nu}_{b}F_{\mu\nu}$. Assuming that $\mathcal{S}$ and $\mathcal{P}$ do not vanish simultaneously, we can block-diagonalize the electromagnetic field tensor through a local Lorentz transformation,
\begin{align}\label{block-diagonalization}
    F_{ab}=
    \begin{pmatrix}
    0&\sigma_1&0&0\\
    -\sigma_1&0&0&0\\
    0&0&0&\sigma_2\\
    0&0&-\sigma_2&0
    \end{pmatrix}.
\end{align}
The Lagrangian of the ModMax theory can be written in terms of $(\sigma_1,\sigma_2)$,
\begin{align}\label{ModMax Lagrangian in the Eigenvalue form}
    \mathcal{B}^{(\gamma)}_{\text{MM}}=\frac{1}{2}(e^{\gamma}\sigma_1^2-e^{-\gamma}\sigma_2^2).
\end{align}
In the following, we formulate the gravitational action associated with the ModMax theory. Assuming that the gravitational Lagrangian $\mathcal{A}^{(\lambda)}$ is a local function of $F_2$, $F_4$, and $\mathcal{R}_n$ for $n=1,2,3,4$. The metric EOM is given by eq.~(\ref{Eigenvalue method matric EoM}). Consider the matrix $\mathcal{R}^{\mu}{}_{\nu}$ can be diagonalized as
\begin{align}
    \mathcal{R}^{\mu}{}_{\nu}=f^{\mu}{}_{a}\mathcal{R}^{a}{}_{b}(f^{-1})^{b}{}_{\nu},\ \ \ \text{with }\mathcal{R}^{a}{}_{b}=\text{diag}(\tilde r_1,\tilde r_2,\tilde r_3,\tilde r_4).
\end{align}
Plugging this expression into the metric EOM, we find
\begin{align}
    &\quad(e^{-1}f)\text{diag}(2\tilde r_1\mathcal{A}^{(\lambda)}_{\tilde r_1},2\tilde r_2\mathcal{A}^{(\lambda)}_{\tilde r_2},2\tilde r_3\mathcal{A}^{(\lambda)}_{\tilde r_3},2\tilde r_4\mathcal{A}^{(\lambda)}_{\tilde r_4})(e^{-1}f)^{-1}\notag\\
    &=\text{diag}(\mathcal{A}^{(\gamma)}-\sigma_1\mathcal{A}^{(\gamma)}_{\sigma_1},\mathcal{A}^{(\gamma)}-\sigma_1\mathcal{A}^{(\gamma)}_{\sigma_1},\mathcal{A}^{(\gamma)}-\sigma_2\mathcal{A}^{(\gamma)}_{\sigma_2},\mathcal{A}^{(\gamma)}-\sigma_2\mathcal{A}^{(\gamma)}_{\sigma_2})
\end{align}
Once again, we can relabel the eigenvalues of $\mathcal{R}^{\mu}{}_{\nu}$ as $\lbrace{r_j}\rbrace$, which satisfy the differential equations
\begin{align}\label{ModMax Eigenvalue method metric EoM}
&2r_{1,2}\mathcal{A}^{(\gamma)}_{r_{1,2}}+\sigma_1\mathcal{A}^{(\gamma)}_{\sigma_1}-\mathcal{A}^{(\gamma)}=0,\notag\\
&2r_{3,4}\mathcal{A}^{(\gamma)}_{r_{3,4}}+\sigma_2\mathcal{A}^{(\gamma)}_{\sigma_2}-\mathcal{A}^{(\gamma)}=0.
\end{align}
Various classes of gravitational actions are capable of reproducing the ModMax action through the metric EOM (\ref{ModMax Eigenvalue method metric EoM}). The simplest of these is given by
\begin{align}\label{general form of A for ModMax}
    \mathcal{A}^{(\gamma)}=a(\gamma)(\sigma_1^2-\sigma_2^2)+\sum_{j=1}^4b_j(\gamma)r_j^{q_j(\gamma)}.
\end{align}
From the metric EOM, we have
\begin{align}
    b_{1,2}r_{1,2}^{q_{1,2}}&=\frac{-a(1-q_3^{-1}-q_4^{-1})(\sigma_1^2+\sigma_2^2)}{q_{1,2}(2-\sum_{j=1}^4q_j^{-1})},\notag\\
    b_{3,4}r^{q_{3,4}}_{3,4}&=\frac{-a(1-q_1^{-1}-q_2^{-1})(\sigma_1^2+\sigma_2^2)}{q_{3,4}(2-\sum_{j=1}^4q_j^{-1})}.
\end{align}
One can plug these equations into (\ref{general form of A for ModMax}) to obtain the effective deformed Lagrangian $\mathcal{B}^{(\gamma)}$. To recover the ModMax Lagrangian (\ref{ModMax Lagrangian in the Eigenvalue form}), the parameters in (\ref{general form of A for ModMax}) should satisfy the constrains
\begin{align}
    \frac{1-q_1^{-1}-q_2^{-1}}{1-q_3^{-1}-q_4^{-1}}=e^{2\gamma},\ \ \ a=\frac{2-\sum_{j=1}^4q_j^{-1}}{4(1-q_1^{-1}-q_2^{-1})}e^{\gamma}.
\end{align}

\subsection{$T\bar T$ deformation}
The $T\bar{T}$ deformation of Maxwell's theory in four dimensions yields the Born-Infeld theory \cite{Conti:2022egv, Ferko:2022iru, Ferko:2023wyi, Ferko:2023ruw}, with the Lagrangian formulated as
\begin{equation}\label{TTbar deformed Maxwell's theory}
	\mathcal{B}^{(\lambda)}_{\mathrm{BI}}=\frac{1}{\lambda}\left(1-\sqrt{1-2\lambda\mathcal{S}-\lambda^2\mathcal{P}^2}\right).
\end{equation}
\subsubsection{Non-minimal $f(\mathcal{R})$ gravity}
The non-minimal $f(\mathcal{R})$ gravitational action corresponding to the Born-Infeld theory can be formally expressed as (\ref{Non-minimal gravity for deformed Maxwell}). Again, an additional parameter $\epsilon$ is introduced in the gravitational action, which reproduces the Born-Infeld theory at its leading order in $\epsilon$.
The gravitational Lagrangian can be specifically constructed as
\begin{align}
\mathcal{A}^{(\lambda)}&=\mathcal{S}+|b(\lambda)|l^{2\epsilon}\mathcal{R}^{2+\epsilon}\Big(2\sqrt{2}\epsilon^{-\frac{1}{2}}(2+\epsilon)-8\lambda|b(\lambda)|\mathcal{R}^2\Big)\notag\\
&\quad+l^{2\epsilon}\mathcal{S}\mathcal{R}^{\epsilon}\Big((\sqrt{2}-1)(1+\frac{\epsilon}{2})-4\lambda|b(\lambda)|\epsilon^{\frac{1}{2}}\mathcal{R}^2\Big)-\frac{\lambda}{4}\epsilon l^{2\epsilon}(2\mathcal{S}^2+\mathcal{P}^2)\mathcal{R}^{\epsilon},
\end{align}
whose structure differs from the scalar case merely by some coefficients and powers. Here, $b(\lambda)$ is an arbitrary function with the correct physical dimension. From the metric EOM (\ref{metric EoM for ModMax}), we have
\begin{align}
    \mathcal{R}^2=\frac{1-\sqrt{2}\lambda\mathcal{S}-\sqrt{1-2\lambda\mathcal{S}-\lambda^2\mathcal{P}^2}}{4\sqrt{2}\lambda\epsilon^{-\frac{1}{2}}|b(\lambda)|}.
\end{align}
Plugging this expression into $\mathcal{A}^{(\lambda)}$, and expanding it in powers of $\epsilon$, we can reproduce the Born-Infeld Lagrangian,
\begin{align}
    \mathcal{A}^{(\lambda)}|_{g=g^{(\lambda)}}&=\frac{1-\sqrt{1-2\lambda\mathcal{S}-\lambda^2\mathcal{P}^2}}{\lambda}+O(\epsilon).
\end{align}

\subsubsection{Eigenvalue method}
The eigenvalue method offers a more systematic framework for constructing the gravitational action associated with Born-Infeld theory, compared with the approach based on non-minimally coupled $f(\mathcal{R})$ gravity. By employing the block-diagonalization (\ref{block-diagonalization}), the Born-Infeld Lagrangian is simplified to
\begin{align}
\mathcal{B}^{(\lambda)}_{\mathrm{BI}}=\frac{1}{\lambda}-\frac{1}{\lambda}\sqrt{(1-\lambda\sigma_1^2)(1+\lambda\sigma_2^2)}.
\end{align}
The second term in the above expression exhibits the same structure as other $T\bar T$ deformation theories. Following the method in subsection \ref{subsection Eigenvalue method}, this part corresponds to the $\tilde{\mathcal{A}}^{(\lambda)}$ term in the gravitational Lagrangian,
\begin{align}
    \tilde{\mathcal{A}}^{(\lambda)}&=-\frac{1}{\lambda}\Big(\sqrt{1-q_1^{-1}-q_2^{-1}+a_1 r_1^{q_1}+a_2 r_2^{q_2}+b_1\sigma_1^{2/(q_1^{-1}+q_2^{-1})}-\lambda\sigma_1^2}\Big)\notag\\
    &\quad\times\Big(\sqrt{1-q_3^{-1}-q_4^{-1}+a_3 r_3^{q_3}+a_4 r_4^{q_4}+b_2\sigma_2^{2/(q_3^{-1}+q_4^{-1})}+\lambda\sigma_2^2}\Big).
\end{align}
By employing the metric EOM (\ref{ModMax Eigenvalue method metric EoM}), we find
\begin{align}
    a_{1,2}r_{1,2}^{q_{1,2}}&=\frac{1-(q_1^{-1}+q_2^{-1})^{-1}b_1\sigma_1^{2/(q_1^{-1}+q_2^{-1})}}{q_{1,2}},\notag\\
    a_{3,4}r_{3,4}^{q_{3,4}}&=\frac{1-(q_3^{-1}+q_4^{-1})^{-1}b_2\sigma_2^{2/(q_3^{-1}+q_4^{-1})}}{q_{3,4}}.
\end{align}
Plugging these equations into $\tilde{\mathcal{A}}^{(\lambda)}$, we have
\begin{align}
    \tilde{\mathcal{A}}^{(\lambda)}|_{g=g^{(\lambda)}}=-\frac{1}{\lambda}\sqrt{(1-\lambda\sigma_1^2)(1+\lambda\sigma_2^2)}.
\end{align}
The remaining constant term $1/\lambda$ in the Born-Infeld Lagrangian should arise from the redundant part $\mathcal{A}^{(\lambda)}_0$ in the gravitational Lagrangian. Since the latter should not alter the metric EOM, it necessarily takes the form
\begin{align}
    \mathcal{A}^{(\lambda)}_0=\Big(\prod_{j=1}^4r_j^{1/2}\Big)F(u_1,u_2),
\end{align}
with $u_1=\sigma_1/\sqrt{r_1r_2}$, and $u_2=\sigma_2/\sqrt{r_3r_4}$. It follows that
\begin{align}
    F(u_1,u_2)=\frac{1}{\lambda}\Big(\prod_{j=1}^{d}f_{j}(u_1,u_2)\Big)^{-1/2},
\end{align}
where the functions $f_j$ are obtained from the following equations,
\begin{align}
    a_{1,2}f_{1,2}^{q_{1,2}}&=\frac{1-(q_1^{-1}+q_2^{-1})^{-1}b_1(u_1\sqrt{f_1f_2})^{2/(q_1^{-1}+q_2^{-1})}}{q_{1,2}},\notag\\
    a_{3,4}f_{3,4}^{q_{3,4}}&=\frac{1-(q_3^{-1}+q_4^{-1})^{-1}b_2(u_2\sqrt{f_3f_4})^{2/(q_3^{-1}+q_4^{-1})}}{q_{3,4}}.
\end{align}
For instance, if we set $q_1=q_2=q_3=q_4=1$, and $a_j\geq0$, the functions $f_j$ take the forms
\begin{align}
    f_{1,2}=\Big(a_{1,2}+\frac{\theta_1}{2}\sqrt{\frac{a_{1,2}}{a_{2,1}}}b_1 u_1\Big)^{-1},\ \ f_{3,4}=\Big(a_{3,4}+\frac{\theta_2}{2}\sqrt{\frac{a_{3,4}}{a_{4,3}}}b_2 u_2\Big)^{-1},
\end{align}
where $\theta_{1}$ and $\theta_2$ are the signs of $(2-b_1\sigma_1)$ and $(2-b_2\sigma_2)$, respectively. The gravitational Lagrangian is explicitly formulated as
\begin{align}
    \mathcal{A}^{(\lambda)}&=\frac{1}{4\lambda}\sqrt{(2\sqrt{a_1a_2r_1r_2}+b_1\theta_1\sigma_1)^2(2\sqrt{a_3a_4r_3r_4}+b_2\theta_2\sigma_2)^2}\notag\\
    &\quad-\frac{1}{\lambda}\sqrt{(a_1 r_1+a_2 r_2+b_1\sigma_1-\lambda\sigma_1^2-1)(a_3 r_3+a_4 r_4+b_2\sigma_2-\lambda\sigma_2^2-1)}.
\end{align}
One can further plug it into the metric EOM and reproduce the Born-Infeld Lagrangian.

\section{Classical-quantum consistency}\label{sec:quantum}
In previous sections, we examined the non-local stress-tensor deformation on a fixed background and the local stress-tensor deformation on a deformed background, and provided their corresponding gravitational theories. This geometric description holds at the classical level: by taking the saddle point with respect to the metric and an independent connection, the gravitational action reduces to the stress-tensor-deformed field-theory action on the deformed metric. In this section, we present a preliminary off-shell study of non-local stress-tensor deformation on a deformed metric background. The deformation incorporates non-local operators, and for specific forms of stress-tensor deformations, the metric dependence of these operators can be systematically expressed in terms of geometric invariants. It should therefore be emphasized that in this scenario, the relationship between the deformed field theory and gravity is established not through the on-shell value of the gravitational action, but rather from the emergence of the gravitational action itself from the structure of the non-local operators. In the following, we will discuss the basic idea of emergent gravity and illustrate its realization through a concrete class of stress-tensor deformations.
\subsection{Basic proposal}\label{sec 6.1}
Our discussion in this section focuses on non-local quadratic stress-tensor deformations. The flow equation of the field theory action is formally written as
\begin{align}\label{deformation for general D and T}
    \partial_{\lambda}S^{(\lambda)}[g,\psi]&=\int d^dx\sqrt{g}\ T_1^{(\lambda)}(\mathcal{G}T_2^{(\lambda)})\notag\\
    &=\int d^dxd^dy\sqrt{g(x)g(y)}\ T_1^{(\lambda)}(x)\langle{x}|\mathcal{G}|{y}\rangle T_2^{(\lambda)}(y),
\end{align}
where $T_1$ and $T_2$ are constructed from the components of the stress tensor. The non-local kernel $\mathcal{G}$ can generally be expressed in terms of the heat kernel $e^{-\tau D}$, where $D$ is a linear differential operator. Furthermore, we require $D$ to be an elliptic operator, which guarantees the smoothness of its heat kernel $\langle{x}|e^{-\tau D}|{y}\rangle$ when $\tau>0$. We now investigate the deformed field theory at the quantum level. The first-order correction to the deformed partition function takes the form
\begin{align}\label{first order deformed partition function}
    \mathcal{Z}^{(\lambda)}&=\int \mathcal{D}\psi\Big(1-\lambda\int d^dx\sqrt{g}\ T_1^{(0)}(\mathcal{G}T_2^{(0)})+O(\lambda^2)\Big)e^{-S^{(0)}[g,\psi]}\notag\\
    &=\mathcal{Z}^{(0)}\Big(1-\lambda\int d^dxd^dy\sqrt{g(x)g(y)}\ \langle{x}|\mathcal{G}|{y}\rangle\langle T_1^{(0)}(x)T_2^{(0)}(y)\rangle+O(\lambda^2)\Big).
\end{align}
Typically, the matrix element $\langle x | \mathcal{G} | y \rangle$ admits an integral representation in terms of the heat kernel $\langle{x}|e^{-\tau D}|{y}\rangle$. The diagonal part of the latter admits a small-$\tau$ asymptotic expansion \cite{DeWitt:1964mxt,Barvinsky:1985an,Vassilevich:2003xt},
\begin{align}\label{Heat kernel expansion formally}
    \langle{x}|e^{-\tau D}|{x}\rangle\sim(4\pi\tau)^{-d/2}\sum_{n=0}^{\infty}a_n(x)\tau^{n/2},
\end{align}
where $\lbrace{a_n(x)}\rbrace$ are known as the Seeley-DeWitt coefficients \cite{Seeley:1967ea,DeWitt:1964mxt} which can be expressed in terms of geometric invariants. To extract the emergent local gravitational action from the deformed partition function (\ref{first order deformed partition function}) using this expansion, we need to appropriately construct the forms of $T_1$ and $T_2$ such that the two-point function $\langle{T_1^{(0)}(x)T_2^{(0)}(y)}\rangle$ contains only contact terms. The simplest way to achieve this is to utilize the $\langle{\Theta\Theta}\rangle$ two-point function derived from the stress tensor Weyl anomaly, which will be discussed in detail in the next subsection.\par
Following the above discussion, we examine the stress-tensor deformation with an inserted linear differential operator $\mathcal{G}=D^{-s}$, and calculate the corresponding emergent local gravitational action. When $D$ is self-adjoint and strictly positive definite, the Schwartz kernel $\langle{x}|D^{-s}|{y}\rangle$ can be expressed via the Mellin transformation \cite{Barvinsky:1985an,berline2003heat},
\begin{align}\label{Mellin transformation}
\langle{x}|D^{-s}|{y}\rangle=\frac{1}{\Gamma(s)}\int_{0}^{\infty}d\tau\ \tau^{s-1}\langle{x}|e^{-\tau D}|{y}\rangle.
\end{align}
The above integral is well-defined over the off-diagonal region $x\neq y$ for any value of $s$. For the diagonal part $x=y$, the integral converges as $\tau\to 0^+$ only if $\Re{s}>d/2$, according to the heat kernel expansion (\ref{Heat kernel expansion formally}). For $\Re s\leq d/2$, regularization is required to yield a finite value for the integral, as will be addressed in the subsequent example.
\subsection{Non-local $\Theta\Theta$ deformation}
Consider the case where the seed theory for the deformation is a conformal field theory, $S^{(0)}=S_{\text{CFT}}$. The stress tensor two-point function can be divided into the connected part and the disconnected part,
\begin{align}
    \langle{T_1^{(0)}(x)T_2^{(0)}(y)}\rangle=\langle{T_1^{(0)}(x)T_2^{(0)}(y)}\rangle_{\text{con}}+\langle{T_1^{(0)}(x)}\rangle\langle{T_2^{(0)}(y)}\rangle.
\end{align}
Contact terms can only arise from the connected part of the two-point function. In order to eliminate the disconnected part, we introduce new operators
\begin{align}
    \tilde T_1^{(0)}=T_1^{(0)}-\langle{T_1^{(0)}}\rangle\mathbf{1},\ \ \tilde T_2^{(0)}=T_2^{(0)}-\langle{T_2^{(0)}}\rangle\mathbf{1},
\end{align}
which satisfy
\begin{align}
    \langle{\tilde T_1^{(0)}(x)\tilde T_2^{(0)}(y)}\rangle=\langle{T_1^{(0)}(x)T_2^{(0)}(y)}\rangle_{\text{con}}.
\end{align}
The simplest choice for $T^{(0)}_1$ and $T^{(0)}_2$ is the trace operator, $\Theta=g^{\mu\nu}T^{(0)}_{\mu\nu}$. In even-dimensional spacetimes, the one-point function $\langle{\Theta}\rangle$ is non-zero, which is known as the Weyl (trace) anomaly. The two-point function can be explicitly calculated using the variational principle,
\begin{align}
    \langle{\Theta(x)\Theta(y)}\rangle_{\text{con}}=\frac{-2}{\sqrt{g(x)g(y)}}g^{\mu\nu}(y)\frac{\delta}{\delta g^{\mu\nu}(y)}\Big(\sqrt{g(x)}\langle{\Theta(x)}\rangle\Big).
\end{align}\par
Let us now consider the two-dimensional case. The Weyl anomaly in two dimensions is given by
\begin{align}
    \langle{\Theta}\rangle_{d=2}=\frac{1}{24\pi}(cR+b\Lambda^2),
\end{align}
where $c$ is the central charge, $b$ is a dimensionless constant, and $\Lambda$ is the UV cutoff. The two-point function $\langle{\Theta\Theta}\rangle$ is given by \cite{Hartman:2023qdn,Hartman:2023ccw},
\begin{align}
    \langle{\Theta(x)\Theta(y)}\rangle_{\text{con}}=\frac{1}{12\pi}\Big(-c\nabla^2_x+b\Lambda^2\Big)\delta(x-y).
\end{align}
We take the differential operator in (\ref{first order deformed partition function}) to be $D=-\nabla^2+\frac{b\Lambda^2}{c}$, and stress tensor components in (\ref{first order deformed partition function}) to be $T_1^{(0)}=T_2^{(0)}=\tilde\Theta$. When $b/c>0$, $D$ is self-adjoint and strictly positive definite. The first-order correction of the partition function takes the form
\begin{align}
    \mathcal{Z}^{(1)}(\mathcal{G}=e^{-\tau D})&=-\frac{\lambda c\mathcal{Z}^{(0)}}{12\pi }\int d^2xd^2y\sqrt{g(x)g(y)}\ \langle{x}|e^{-\tau D}|{y}\rangle D_x\delta(x-y)\notag\\
    &=\frac{\lambda c\mathcal{Z}^{(0)}}{12\pi}\int d^2x\sqrt{g}\ \partial_{\tau}\langle{x}|e^{-\tau D}|{x}\rangle, 
\end{align}
where we have used the identity $(\partial_{\tau}+D_{x})\langle{x}|e^{-\tau D}|{y}\rangle=0$. By employing the heat kernel expansion of $D$, we find
\begin{align}
    &\quad \mathcal{Z}^{(1)}(\mathcal{G}=e^{-\tau D})\notag\\
    &=\frac{\lambda c\mathcal{Z}^{(0)}}{48\pi^2}\int d^2x\sqrt{g}\Big(\sum_{n=0}^{\infty}(n-1)a_{2n}\tau^{n-2}\Big)\notag\\
    &=\frac{\lambda c\mathcal{Z}^{(0)}}{48\pi^2}\int d^2x\sqrt{g}\Big(\frac{-1}{\tau^2}+\frac{b^2\Lambda^4}{4c^2}+\frac{b\Lambda^2}{12c}R+\frac{1}{60}\nabla^2R+\frac{1}{120}R^2+O(\tau)\Big).
\end{align}
Next, we compute the first-order correction to the partition function induced by the deformation $\tilde\Theta D^{-s}\tilde\Theta$. By employing the Mellin transformation (\ref{Mellin transformation}), we have
\begin{align}\label{Z1 D^-s}
    \mathcal{Z}^{(1)}(\mathcal{G}=D^{-s})&=\frac{\lambda c\mathcal{Z}^{(0)}}{12\pi \Gamma(s)}\int d^2x\sqrt{g}\int_{0}^{\infty}d\tau\ \tau^{s-1}\partial_{\tau}\langle{x}|e^{-\tau D}|{x}\rangle.
\end{align}
It is evident that for $\Re s>2$, the integral over $\tau\in (0,\infty)$ converges at both limits. By introducing a new set of expansion coefficients $\lbrace A_{2n}(x)\rbrace$ defined as
\begin{align}
     \langle{x}|e^{-\tau D}|{x}\rangle=\frac{1}{4\pi\tau}e^{-\frac{b\Lambda^2\tau}{c}}(A_0(x)+\tau A_2(x)+\tau^2A_{4}(x)+...),
\end{align}
The first-order correction can be written as
\begin{align}
    \mathcal{Z}^{(1)}(\mathcal{G}=D^{-s})&=\frac{\lambda c\mathcal{Z}^{(0)}}{48\pi^2 \Gamma(s)}\int d^2x\sqrt{g}\Big[-\Big(\frac{c}{b\Lambda^2}\Big)^{s-2}\Gamma(s-2)A_0\notag\\
    &\quad+\sum_{n=0}^{\infty}\Big(\frac{c}{b\Lambda^2}\Big)^{n+s-1}\Gamma(n+s-1)(nA_{2n+2}-\frac{b\Lambda^2}{c}A_{2n})\Big].
\end{align}
In the regime $\Re s\leq 2$, the integral (\ref{Z1 D^-s}) exhibits UV divergence as $\tau\to 0^+$. We employ the proper-time regularization \cite{Vassilevich:2003xt} by introducing a cutoff at $\tau=\epsilon$,
\begin{align}
    \mathcal{Z}^{(1)}_{\text{reg}}(\mathcal{G}=D^{-s})&=\frac{\lambda c\mathcal{Z}^{(0)}}{12\pi \Gamma(s)}\int d^2x\sqrt{g}\int_{\epsilon}^{\infty}d\tau\ \tau^{s-1}\partial_{\tau}\langle{x}|e^{-\tau D}|{x}\rangle\notag\\
    &=\frac{\lambda c\mathcal{Z}^{(0)}}{48\pi^2 \Gamma(s)}\int d^2x\sqrt{g}\sum_{n=-2}^{\lfloor{-\Re s}\rfloor}\sum_{k=0}^{\lfloor{-\Re s-n}\rfloor}c_n\Big[(1-\delta_{k+n+s})\frac{(-\frac{b\Lambda^2}{c})^k}{k!(k+n+s)}\epsilon^{k+n+s}\notag\\
    &\quad+\delta_{k+n+s}\frac{(-\frac{b\Lambda^2}{c})^{k}}{k!}\ln\epsilon+O(\epsilon^0)\Big],
\end{align}
where $c_{-2}=A_0$, $c_{n>-2}=(\frac{b\Lambda^2}{c})A_{2n+2}-(n+1)A_{2n+4}$, $\delta_{0}=1$, and $\delta_{n\neq 0}=0$. Only the expansion coefficients $\lbrace{A_{2j}|j=0,1,...,\lfloor{-\Re s}\rfloor+2}\rbrace$ contribute to the divergent part of the integral, whereas the remaining coefficients give rise to its finite part in the limit $\epsilon\to 0$.\par
The approach for calculating the first-order deformed partition function under the $(\tilde{\Theta} e^{-\tau D} \tilde{\Theta})$ deformation can be extended to higher dimensions. In such cases, the local operator $D$ should be selected as an appropriate higher-order differential operator, depending on the specific form of the two-point function $\langle{\Theta\Theta}\rangle_{\text{con}}$. In four dimensions, for instance, the connected two-point function can be derived from the four-dimensional Weyl anomaly \cite{Hartman:2023ccw, Hartman:2023qdn},
\begin{align}
    \langle{\Theta(x)\Theta(y)}\rangle_{\text{con}}&=-6b_1\Big(\nabla^4+C^{\mu\nu}\nabla_{\mu}\nabla_{\nu}+H^{\mu}\nabla_{\mu}+P\Big)\delta(x-y),
\end{align}
where $C^{\mu\nu}=(4a/3b_1)G^{\mu\nu}+(R/3+b_2\Lambda^2/b_1)g^{\mu\nu}$, $H^{\mu}=(1/3)\nabla^{\mu}R$, and $P=-(b_2\Lambda^2R+2b_3\Lambda^4)/3b_1$. Here $b_1$, $b_2$, and $b_3$ are dimensionless coefficients. The deformation involves a fourth-order differential operator defined as $D=\nabla^4+C^{\mu\nu}\nabla_{\mu}\nabla_{\nu}+E^{\mu}\nabla_{\mu}+F$, with $E^{\mu}=2\nabla_{\nu}C^{\mu\nu}-H^{\mu}$ and $F=\nabla_{\mu}\nabla_{\nu}C^{\mu\nu}-\nabla_{\mu}H^{\mu}+P$. The first-order correction to the partition function is evaluated employing the heat kernel coefficients for such a  fourth-order operator derived in \cite{Barvinsky:2021ijq},
\begin{align}
    \mathcal{Z}^{(1)}(\mathcal{G}=e^{-\tau D})&=-6\lambda b_1\mathcal{Z}^{(0)}\int d^4x\sqrt{g}\ \partial_{\tau}\langle{x}|e^{-\tau D}|{x}\rangle\notag\\
&=-\frac{3\lambda b_1\mathcal{Z}^{(0)}}{8\pi^2}\int d^4x\sqrt{g}\Big[\frac{-1}{2\tau^2}+\frac{1}{4}\Big(\frac{1}{18}R^2-\frac{1}{45}R^{\mu\nu}R_{\mu\nu}\notag\\
    &\quad+\frac{1}{45}R^{\mu\nu\rho\sigma}R_{\mu\nu\rho\sigma}+\frac{2}{15}\nabla^2 R-\frac{1}{3}C^{\mu\nu}R_{\mu\nu}+\frac{1}{6}CR+\frac{1}{24}C^2\notag\\
    &\quad+\frac{1}{12}C^{\mu\nu}C_{\mu\nu}-\frac{5}{9}\nabla_{\mu}\nabla_{\nu}C^{\mu\nu}+\frac{2}{9}\nabla^2 C+\nabla_{\mu}E^{\mu}-2F\Big)+O(\tau)\Big].
    \end{align}
The first-order correction to the partition function with a non-local kernel $\mathcal{G}=D^{-s}$ can be computed using the approach analogous to that in two dimensions.

\subsection{Deformed 4d scalar field theory}
In section \ref{section nonlocal}, we studied Einstein gravity from a classical perspective and found that, by taking the metric saddle point, the linearized gravitational action can be expressed as a coupling between the seed theory and a non-local stress-energy tensor deformation term. In this section, we aim to recover Einstein gravity from a quantum perspective (at leading order). To achieve this, the non-local deformation should be introduced on a dynamical background (without taking the classical saddle point), and then recast, employing the formalism of subsection \ref{sec 6.1}, into an expansion in terms of geometric invariants. For simplicity, we adopt the leading-order deformation obtained in Eq.~(\ref{first-order deformed action}) to perturb the seed theory on the dynamical background $g$,
\begin{align}
    S^{(\lambda)}=S_0+\frac{\lambda}{2l^2}\!\int\!d^dxd^dy\sqrt{g(x)g(y)}T^{\mu\nu}(x)\mathcal{G}_{\mu\nu\rho\sigma}(x,y)T^{\rho\sigma}(y)+O(\lambda^2),
\end{align}
where $\mathcal{G}_{\mu\nu\rho\sigma}(x,y)=G_{\mu\nu\rho\sigma}(x,y)-1/(d-2)G_{\mu\nu\alpha\beta}(x,y)g^{\alpha\beta}(y)g_{\rho\sigma}(y)$ and the Green's function satisfies $[-g^{\mu\rho}g^{\nu\sigma}\Box-R^{\mu\rho\nu\sigma}-R^{\nu\rho\mu\sigma}]_xG_{\rho\sigma\alpha\beta}(x,y)=\delta(x-y)\delta^{\mu}_{\alpha}\delta^{\nu}_{\beta}$. One may choose different forms for the non-local kernel $\mathcal{G}$ at leading order. The subsequent analysis shows that, with suitable renormalization schemes, the leading-order gravitational action can always be recast as Einstein gravity. Different kernels, however, yield distinct structures in the higher-derivative gravitational sector.\par
In the following analysis, we take the four-dimensional free massive scalar field theory as the seed theory, $S_0=\frac{1}{2}\int d^4x\sqrt{g}\phi(-\Box+m^2)\phi$. The one-loop effective action \cite{Vassilevich:2003xt} is given by
\begin{align}
    W_0[g]=\frac{1}{2}\text{ln }\text{det}(-\Box+m^2).
\end{align}
Using the zeta-function regularization, the functional determinant can be expressed in terms of the heat kernel of the operator $D=(-\Box+m^2)$,
\begin{align}
    W_0[g]=-\frac{1}{2}\int^{\infty}_{0}\frac{d\tau}{\tau}\text{tr}(e^{-\tau D}).
\end{align}
For $m \neq 0$, the integral converges in the infrared limit $\tau \to \infty$; its ultraviolet divergence as $\tau \to 0^+$ is governed by the leading terms of the heat‑kernel expansion. We employ the proper-time regularization by introducing the UV cutoff at $\tau=\Lambda^{-2}$. From the heat kernel expansion, the regularized one-loop effective action is given by
\begin{align}
    W^{(\Lambda)}_{0}[g]&=-\frac{1}{2}\int^{\infty}_{\Lambda^{-2}}\frac{d\tau}{\tau}\text{tr}(e^{-\tau D})\notag\\
    &=-\frac{1}{64\pi^2}\int d^4x\sqrt{g}\Big[\Lambda^4+\frac{\Lambda^2}{3}\Big(R-6m^2\Big)-\frac{2\text{ln}(\Lambda/m)}{3}\Big(m^2R-3m^4-\frac{1}{5}\Box R\notag\\
    &\quad-\frac{1}{12}R^2+\frac{1}{30}R_{\mu\nu}R^{\mu\nu}-\frac{1}{30}R_{\mu\nu\rho\sigma}R^{\mu\nu\rho\sigma}\Big)\Big]+O(\Lambda^0).
\end{align}
The stress-tensor two-point function can be calculated by the variational principle
\begin{align}\label{2pt general}
    \langle{T^{\mu\nu}(x)T^{\rho\sigma}(y)}\rangle^{(\Lambda)}_0&=\frac{4}{\sqrt{g(x)}\sqrt{g(y)}}\frac{\delta^2 W_{\Lambda}}{\delta g_{\mu\nu}(x)\delta g_{\rho\sigma}(y)}\notag\\
    &=-\frac{1}{64\pi^2}\Big[(\Lambda^4-2m^2\Lambda^2+2m^4\text{ln}(\Lambda/m))(g^{\mu\nu}g^{\rho\sigma}-g^{\mu\rho}g^{\nu\sigma}-g^{\mu\sigma}g^{\nu\rho})\delta(x-y)\notag\\
    &\quad+\frac{2}{3}(\Lambda^2-2m^2\text{ln}(\Lambda/m))\Big((2g^{\mu\nu}g^{\rho\sigma}+g^{\mu\rho}g^{\nu\sigma}+g^{\mu\sigma}g^{\nu\rho})\Box-2g^{\mu\nu}\nabla^{\rho}\nabla^{\sigma}\notag\\
    &\quad-2g^{\rho\sigma}\nabla^{\mu}\nabla^{\nu}-4R^{\mu\rho\nu\sigma}+2g^{\mu\nu}R^{\rho\sigma}+2g^{\rho\sigma}R^{\mu\nu}-g^{\mu\rho}R^{\nu\sigma}-g^{\mu\sigma}R^{\nu\rho}\notag\\
    &\quad-g^{\nu\rho}R^{\mu\sigma}-g^{\nu\sigma}R^{\mu\rho}\Big)\delta(x-y)+\text{ln}(\Lambda/m)\mathcal{F}_{(4)}(x,y)\Big]+O(\Lambda^0),
\end{align}
where $\mathcal{F}_{(4)}(x,y)$ involves fourth-order derivatives of the $\delta$-function, including $\nabla^4\delta$, $R\nabla^2\delta$, $R^2\delta$, etc.\par
Next, we investigate the UV expansion of the Green's function $G_{\mu\nu\rho\sigma}$, which can be represented via the heat kernel of the corresponding Laplacian-type operator
,
\begin{align}
    G_{\mu\nu\rho\sigma}^{(\Lambda)}(x,y)=\int_{\Lambda^{-2}}^{\infty}d\tau\ K_{\mu\nu\rho\sigma}(\tau;x,y).
\end{align}
In four dimensions, the heat kernel has the following small-$\tau$ expansion,
\begin{align}\label{off-diag K}
    K_{\mu\nu\rho\sigma}(\tau;x,y)=\frac{\Delta^{1/2}(x,y)}{16\pi^2\tau^2}e^{-\sigma(x,y)/2\tau}\sum_{n=0}^{\infty}a^{(2n)}_{\mu\nu\rho\sigma}(x,y)\tau^n,
\end{align}
where $\sigma$ is the Synge world function and $\Delta$ is the Van Vleck–Morette determinant. The trace of the heat kernel can be written in terms of the geometric invariants,
\begin{align}
     K_{\mu\nu\rho\sigma}(\tau;x,x)&=\frac{1}{16\pi^2}\Big[\frac{1}{2\tau^2}(g_{\mu\rho}g_{\nu\sigma}+g_{\mu\sigma}g_{\nu\rho})-\frac{1}{\tau}\Big(R_{\mu\rho\nu\sigma}+R_{\nu\rho\mu\sigma}\notag\\
     &\quad-\frac{1}{12}R(g_{\mu\rho}g_{\nu\sigma}+g_{\mu\sigma}g_{\nu\rho})\Big)+O(\tau^0)\Big].
\end{align}
By combining the stress-tensor two-point function (\ref{2pt general}) and the heat kernel (\ref{off-diag K}), and employing the coincidence limit $x\to y$ of derivatives of the Synge world function and the Van Vleck–Morette determinant \cite{Avramidi:2000bm, Avramidi:2001ns, Vassilevich:2003xt}, we compute the first-order correction to the partition function (\ref{first order deformed partition function}) in a systematic expansion,
\begin{align}\label{eq162}
\mathcal{Z}^{(\lambda)}_{\text{reg}}&=\mathcal{Z}_0+\frac{\lambda\mathcal{Z}_0}{128\pi^2l^2}\int d^4x\sqrt{g}\Big[-\frac{1}{4\pi^2}\Big(13\Lambda^6-10m^2\Lambda^4-16m^2\Lambda^4\text{ln}(\Lambda/m)+10m^4\Lambda^2\text{ln}(\Lambda/m)\Big)\notag\\
&\quad-\frac{R}{12\pi^2}\Big(7\Lambda^4+8\Lambda^4\text{ln}(\Lambda/m)-14m^2\Lambda^2\text{ln}(\Lambda/m)+16m^4(\text{ln}(\Lambda/m))^2\Big)+O(\nabla^4)\Big]\notag\\
&=\mathcal{Z}_0+\frac{\lambda\mathcal{Z}_0}{2l^2}\int d^{4}x\sqrt{g}\,
\Big[
\alpha_0+\alpha_{1} R
+O(\nabla^4)
\Big].
\end{align}
where $O(\nabla^4)$ represents the contribution of higher-derivative gravitational terms.\par
In the following, we demonstrate that Einstein gravity, as given in Eq.~(\ref{Einstein gravity}), can be reproduced at the leading order of $Z_{\rm reg}^{(\lambda)}$. This requires specifying a
renormalization prescription that fixes (i) the vacuum-energy term and (ii) the curvature term at a
chosen reference scale, namely scheme I. In the proper-time regularization, imposing the condition that
the induced vacuum-energy contribution vanishes, $\alpha_0=0$, yields a relation $m=\beta\Lambda$,
where the dimensionless constant $\beta$ is determined by
\begin{align}\label{A79}
    13-10\beta^2+16\beta^2\ln\beta-10\beta^4\ln\beta=0.
\end{align}
It can be shown that this equation admits a unique real solution. By matching the coefficient of the curvature term, $\alpha_1$ is fixed and thereby defines the
induced Newton constant at the same reference scale. This matching can be
implemented by choosing $\Lambda=\gamma\sqrt{l^{2}/\lambda}$, where $\gamma$ is another
dimensionless constant satisfying
\begin{align}\label{A80}
    \gamma^4=\frac{768\pi^4}{7-8\ln\beta+14\beta^2\ln\beta+16\beta^4(\ln\beta)^2}.
\end{align}
Near the numerical solution of (\ref{A79}), the right-hand side of Eq.~(\ref{A80}) remains positive, ensuring that $\gamma$ admits a real solution. By putting everything together, the regularized partition function can be rewritten as
\begin{align}\label{eq 165}
    \mathcal{Z}^{(\lambda)}_{\text{reg}}&=\mathcal{Z}_0-\frac{\mathcal{Z}_0l^2}{2\lambda}\int d^4x\sqrt{g}\Big[R+O(\lambda\nabla^4)\Big].
\end{align}
Note that the leading-order term now scales as $\lambda^{-1}$. Since the preceding computation relies on the small-$\lambda$ expansion, maintaining consistency requires choosing the metric $g=\hat\gamma+\lambda h$, where $h$ is the auxiliary field (without taking the classical saddle point), and $\hat\gamma$ is the reference metric satisfying the vacuum Einstein's equation.\par
Next we address the gravitational action within a more rigorous renormalization scheme alternatively, namely scheme II. From equation (\ref{eq162}), the one-loop contribution can be written in the form of an effective action,
\begin{align}
    S_{\text{eff}}&=\frac{\lambda}{128\pi^2l^2}\int d^4x\sqrt{g}\Big[\frac{1}{4\pi^2}\Big(13\Lambda^6-10m^2\Lambda^4-16m^2\Lambda^4\text{ln}(\Lambda/m)+10m^4\Lambda^2\text{ln}(\Lambda/m)\Big)\notag\\
&\quad+\frac{R}{12\pi^2}\Big(7\Lambda^4+8\Lambda^4\text{ln}(\Lambda/m)-14m^2\Lambda^2\text{ln}(\Lambda/m)+16m^4(\text{ln}(\Lambda/m))^2\Big)+O(\nabla^4)\Big].
\end{align}
Next, we introduce the local counterterms to cancel the divergent parts, leading to the renormalized action. We note that both the cosmological-constant term and the Einstein–Hilbert term exhibit power-law divergences in the UV cutoff $\Lambda$, which are removed by appropriate counterterms. In addition, the Einstein–Hilbert term contains a purely logarithmic divergence, $16m^4(\ln(\Lambda/m))^2$. Upon introducing the renormalization scale $\mu$, this logarithmic term can be rewritten as
\begin{align}
    16m^4(\ln(\Lambda/m))^2=16m^4\Big((\ln(\Lambda/\mu))^2+2\ln(\Lambda/\mu)\ln(\mu/m)+(\ln(\mu/m))^2\Big).
\end{align}
The last term has the finite contribution after renormalization. Thus, the renormalized effective action can be written as
\begin{align}
    S_{\text{ren}}(\mu)&=\frac{\lambda}{128\pi^2l^2}\int d^4x\sqrt{g}\Big[\frac{4m^4}{3\pi^2}(\text{ln}(\mu/m))^2R+O(\nabla^4)\Big]\notag\\
    &=\frac{1}{16\pi G_{\text{eff}}(\mu)}\int d^4x\sqrt{g}\Big[R+O(\nabla^4)\Big],
\end{align}
where $G_{\text{eff}}=\frac{6\pi^3l^2}{\lambda m^4}(\ln(\mu/m))^{-2}$. By imposing the matching condition at the scale $\mu=\mu_0=m\exp(\frac{4\sqrt{3}\pi^2l^2}{\lambda m^2})$, the gravitational action is restored to the form given in (\ref{eq 165}).

The two renormalization prescriptions derive from the same ultraviolet expansion but differ in how local divergences are fixed.
In Scheme I, the proper-time cutoff is retained, and local terms are fixed by matching at a reference scale. The vacuum-energy contribution is set to zero by relating the mass to the cutoff, and the curvature term is normalized by identifying the cutoff with the deformation parameter. This procedure yields a leading local term proportional to $\ell^{2}/\lambda\int\!\sqrt{g}\, R$. Consistency of the small-$\lambda$ expansion requires perturbing around a background solution of the vacuum Einstein equations.

In Scheme II, power divergences are absorbed into local counterterms, and the remaining finite curvature term defines an induced Newton constant $G_{\mathrm{eff}}(\mu)$ at renormalization scale $\mu$. The resulting Einstein--Hilbert term exhibits explicit scale dependence, consistent with effective-field-theory expectations. No relations are imposed between the cutoff and physical parameters beyond standard renormalization conditions.

Both schemes generate an induced Einstein--Hilbert term at leading derivative order. Scheme I implements a direct matching to the Einstein form at a fixed scale, whereas Scheme II preserves regulator independence and renormalization-group structure. In this work, we adopt Scheme II, as it maintains a clear separation between ultraviolet regularization and physical parameters and avoids regulator-dependent relations among couplings. Scheme II may be regarded as a specific matching prescription that makes the emergence of linearized Einstein dynamics explicit.

\section{Conclusion and outlook}\label{sec:conclusion}

In this work, we developed a unified \emph{geometric realization} of stress-tensor deformations, in which a broad class of deformed quantum (or classical) field theories can be encoded by a gravitational functional whose metric path integral is evaluated at a saddle. This perspective provides a concrete bridge between stress-tensor flows and (semi)classical gravitational dynamics: the deformation parameter is promoted to a coupling on the gravity side, while the resulting saddle metric organizes the deformation into a covariant description on the deformed geometry.

A key outcome is a practical framework that treats, on the same footing, (i) intrinsically non-local stress-tensor deformations defined on a fixed reference background and (ii) local (or effectively local) stress-tensor deformations defined on the dynamically deformed background. Technically, we showed that bilocal deformations can be reorganized into a controlled derivative expansion built from geometric invariants, thereby providing a systematic perturbative scheme that is both computable and manifestly covariant. This reorganization also clarifies how different choices of non-local kernels correspond to distinct higher-derivative completions in the emergent gravitational action.

We illustrated the general construction with several explicit and nontrivial examples. For scalar theories, the geometric prescription naturally reproduces Nambu-Goto-type structures and their higher-dimensional generalizations, and it extends straightforwardly to interacting potentials. For multiple scalars, the formalism accommodates non-linear models such as Born-Infeld- and ModMax-like systems. For gauge dynamics, we applied the method to four-dimensional Maxwell theory, 
providing gravitational realizations for stress-tensor deformations that generate non-linear electrodynamics—including root-$T\bar T$-type and $T\bar T$-type flows—via both Palatini $f(R)$ constructions and an eigenvalue-based approach.

Beyond the classical mapping, we addressed a central consistency requirement: whether the same geometric picture can be recovered from the \emph{quantum} perspective when the deformation is introduced on a dynamical background. By combining non-local deformation with heat-kernel/derivative-expansion techniques, we demonstrated (at leading order, with appropriate renormalization prescriptions) that the resulting effective action can be recast as Einstein gravity supplemented by higher-derivative corrections. This establishes a concrete classical-quantum correspondence for the emergent gravitational description and identifies the origin of scheme- and kernel-dependence in subleading sectors.

There are several natural directions to pursue:
\begin{itemize}
  \item \textbf{Higher-order and nonperturbative structure.} Extending the construction beyond leading order---including possible resummations, multi-saddle contributions, and global issues---would sharpen the predictive power of the correspondence and clarify its regime of validity.
  \item \textbf{Constraints from consistency.} It would be important to systematically study unitarity, causality, locality/analyticity constraints, and positivity properties of the induced gravitational effective action, especially in higher dimensions where genuinely non-local effects are unavoidable.
  \item \textbf{Broader matter content and symmetries.} Generalizations to fermions, non-Abelian gauge theories, supersymmetric systems, and theories with anomalies should further test the universality of the geometric realization and may reveal symmetry-protected structures in the deformation space.
  \item \textbf{Observables on curved backgrounds.} Computing correlation functions, entanglement measures, and finite-temperature observables directly in the deformed geometry would provide sharp diagnostics and facilitate comparisons with holography and other emergent-gravity paradigms.
\end{itemize}

Overall, our results establish a flexible and computable geometric framework for stress-tensor deformations that unifies local and non-local flows and makes the emergence of gravitational dynamics from stress-tensor data explicit. We expect this approach to serve as a useful platform for systematically exploring stress-tensor deformations across diverse dimensions and for clarifying the conceptual links among irrelevant deformations, non-locality, and emergent spacetime geometry.

\section*{Acknowledgments}
We thank Ronggen Cai, Bo Feng, Mingzhe Li,  Roberto Tateo, and Stefan Theisen for helpful discussions. We are particularly grateful to Tianhao Wu for informing us to apply the heat-kernel to enable our completely independent implementation of the main-text derivation, whereby a massive scalar field, under a deformation, yields linearized Einstein gravity. This work was supported by NSFC Grant Nos. 12475053, 12475047, 12588101, and 12235016.

\appendix

\section{Second-order non-local deformation induced by Einstein gravity}\label{appendix second order}
Based on the leading-order results of the non-local stress-tensor deformation induced by Einstein gravity calculated in subsection \ref{subsection leading order}, we give the second-order contribution in this appendix.\par
According to the expansion of the Einstein-Hilbert action, one can find the expression of $\tilde{\mathcal{F}}^{(2)\mu\nu}$ in (\ref{full EOM of h}):
\begin{align}
	\tilde{\mathcal{F}}^{(2)\mu\nu}[h]&=\frac{1}{8}\Big(4\hat\nabla_{\rho}h^{\mu\nu}\hat\nabla_{\sigma}h^{\rho\sigma}-8\hat\nabla_{\rho}h^{\mu\rho}\hat\nabla_{\sigma}h^{\nu\sigma}-2h^{\mu\nu}\hat\nabla_{\sigma}\hat\nabla_{\rho}h^{\rho\sigma}+8h^{\nu\rho}\hat\nabla_{\sigma}\hat\nabla_{\rho} h^{\mu\sigma}\nonumber\\
	&\quad+4h^{\rho\sigma}\hat\nabla_{\sigma}\hat\nabla_{\rho}h^{\mu\nu}+2h^{\mu\nu}\hat\Box h^{\rho}_{\rho}-8h^{\nu\rho}\hat\Box h^{\mu}_{\rho}+2h^{\rho}_{\rho}\hat\Box h^{\mu\nu}-8h^{\mu\rho}(\hat\nabla_{\rho}\hat\nabla_{\sigma}h^{\nu\sigma}\notag\\
	&\quad+\hat\nabla_{\sigma}\hat\nabla_{\rho}h^{\nu\sigma}-\hat\Box h^{\nu}_{\rho}-\hat\nabla_{\sigma}\hat\nabla^{\nu}h_{\rho}^{\sigma})-4h^{\rho}_{\rho}\hat\nabla_{\sigma}\hat\nabla^{\nu}h^{\mu\sigma}-\hat\gamma^{\mu\nu}\hat\nabla_{\sigma}h^{\delta}_{\delta}\hat\nabla^{\sigma}h^{\rho}_{\rho}\notag\\
	&\quad+8\hat\nabla_{\sigma}h^{\nu}_{\rho}\hat\nabla^{\sigma}h^{\mu\rho}-4\hat\gamma^{\mu\nu}\hat\nabla_{\rho}h^{\rho\sigma}\hat\nabla_{\delta}h_{\sigma}^{\delta}+4\hat\gamma^{\mu\nu}\hat\nabla^{\sigma}h^{\rho}_{\rho}\hat\nabla_{\delta}h_{\sigma}^{\delta}+2\hat\gamma^{\mu\nu}h^{\rho}_{\rho}\hat\nabla_{\delta}\hat\nabla_{\sigma}h^{\sigma\delta}\nonumber\\
	&\quad-2\hat\gamma^{\mu\nu}h^{\rho}_{\rho}\hat\Box h^{\sigma}_{\sigma}-2\hat\gamma^{\mu\nu}\hat\nabla_{\sigma}h_{\rho\delta}\hat\nabla^{\delta}h^{\rho\sigma}+3\hat\gamma^{\mu\nu}\hat\nabla_{\delta}h_{\rho\sigma}\hat\nabla^{\delta}h^{\rho\sigma}+4\hat\nabla_{\rho}h^{\nu\rho}\hat\nabla^{\mu}h^{\sigma}_{\sigma}\notag\\
	&\quad+4h^{\rho}_{\rho}\hat\nabla^{\mu}\hat\nabla_{\sigma}h^{\nu\sigma}-4h^{\rho\sigma}\hat\nabla^{\mu}\hat\nabla^{\nu}h_{\rho\sigma}-8\hat\nabla^{\sigma}h^{\mu\rho}\hat\nabla^{\nu}h_{\rho\sigma}+4\hat\nabla^{\mu}h^{\rho\sigma}\hat\nabla^{\nu}h_{\rho\sigma}\notag\\
	&\quad+4\hat\nabla_{\rho}h^{\mu\rho}\hat\nabla^{\nu}h^{\sigma}_{\sigma}-2\hat\nabla^{\mu}h^{\rho}_{\rho}\hat\nabla^{\nu}h^{\sigma}_{\sigma}-2\hat\nabla_{\rho}h^{\sigma}_{\sigma}(\hat\nabla^{\rho}h^{\mu\nu}-2\hat\nabla^{\nu}h^{\mu\rho})\nonumber\\
	&\quad-8\hat\nabla_{\sigma}h_{\rho}^{\sigma}\hat\nabla^{\nu}h^{\mu\rho}-4h^{\rho}_{\rho}\hat\nabla^{\nu}\hat\nabla_{\sigma}h^{\mu\sigma}+2h^{\rho}_{\rho}\hat\nabla^{\nu}\hat\nabla^{\mu}h^{\sigma}_{\sigma}\Big).\label{F2}
\end{align}
From the full EOM (\ref{full EOM of h}), we can extract the the $\lambda^2$ order equation
\begin{equation}
	\frac{1}{2}\hat \Box h_{[2]}^{\mu\nu}-\frac{1}{4}\hat\gamma^{\mu\nu}\hat \Box h^{\ \alpha}_{[2]\alpha}+\frac{1}{2}\left(\hat R^{\mu\rho\nu\sigma}+\hat R^{\nu\rho\mu\sigma}\right)h_{[2]\rho\sigma}-\frac{1}{l^2}h_{[1]\rho\sigma}M^{\mu\nu\rho\sigma}+\tilde{\mathcal{F}}^{(2)\mu\nu}[h_{[1]}]=0,
\end{equation}
where
\begin{equation}
	M^{\mu\nu\rho\sigma}=\frac{\partial \hat T^{\mu\nu}}{\partial\hat\gamma_{\rho\sigma}}+\frac{\partial \hat T^{\rho\sigma}}{\partial\hat\gamma_{\mu\nu}}+\frac{1}{2}\hat T^{\mu\nu}\hat\gamma^{\rho\sigma}+\frac{1}{2}\hat T^{\rho\sigma}\hat\gamma^{\mu\nu},
\end{equation}
and $\tilde{\mathcal{F}}^{(2)\mu\nu}[h_{[1]}]$ denotes replacing all $h_{\mu\nu}$ in (\ref{F2}) with its first-order coefficient $h_{[1]\mu\nu}$, which has been obtained in (\ref{saddle point metric at the first order}). By solving this equation, we can determine the saddle-point metric perturbation up to the second order:
\begin{align}
	h^*_{\mu\nu}(x)&=-2\lambda t_{\mu\nu}(x)+4\lambda^2l^{-2}\int d^dy\sqrt{\hat\gamma(y)}G_{\mu\nu\eta\xi}(x,y)\Big(\delta^{\eta}_{\alpha}\delta^{\xi}_{\beta}-\frac{1}{d-2}\hat\gamma^{\eta\xi}(y)\hat\gamma_{\alpha\beta}(y)\Big)\nonumber\\
	&\quad\times\Big[M^{\alpha\beta\rho\sigma}t_{\rho\sigma}+\frac{l^{2}}{4}\Big(4\hat\nabla_{\rho}t^{\alpha\beta}\hat\nabla_{\sigma}t^{\rho\sigma}-8\hat\nabla_{\rho}t^{\alpha\rho}\hat\nabla_{\sigma}t^{\beta\sigma}-2t^{\alpha\beta}\hat\nabla_{\sigma}\hat\nabla_{\rho}t^{\rho\sigma}\notag\\
	&\quad+8t^{\beta\rho}\hat\nabla_{\sigma}\hat\nabla_{\rho} t^{\alpha\sigma}+4t^{\rho\sigma}\hat\nabla_{\sigma}\hat\nabla_{\rho}t^{\alpha\beta}+2t^{\alpha\beta}\hat\Box t^{\rho}_{\rho}-8t^{\beta\rho}\hat\Box t^{\alpha}_{\rho}+2t^{\rho}_{\rho}\hat\Box t^{\alpha\beta}\notag\\
	&\quad-8t^{\alpha\rho}(\hat\nabla_{\rho}\hat\nabla_{\sigma}t^{\beta\sigma}+\hat\nabla_{\sigma}\hat\nabla_{\rho}t^{\beta\sigma}-\hat\Box t^{\beta}_{\rho}-\hat\nabla_{\sigma}\hat\nabla^{\beta}t_{\rho}^{\sigma})-4t^{\rho}_{\rho}\hat\nabla_{\sigma}\hat\nabla^{\beta}t^{\alpha\sigma}\nonumber\\
	&\quad-\hat\gamma^{\alpha\beta}\hat\nabla_{\sigma}t^{\delta}_{\delta}\hat\nabla^{\sigma}t^{\rho}_{\rho}+8\hat\nabla_{\sigma}t^{\beta}_{\rho}\hat\nabla^{\sigma}t^{\alpha\rho}-4\hat\gamma^{\alpha\beta}\hat\nabla_{\rho}t^{\rho\sigma}\hat\nabla_{\delta}t_{\sigma}^{\delta}+4\hat\gamma^{\alpha\beta}\hat\nabla^{\sigma}t^{\rho}_{\rho}\hat\nabla_{\delta}t_{\sigma}^{\delta}\notag\\
	&\quad+2\hat\gamma^{\alpha\beta}t^{\rho}_{\rho}\hat\nabla_{\delta}\hat\nabla_{\sigma}t^{\sigma\delta}-2\hat\gamma^{\alpha\beta}t^{\rho}_{\rho}\hat\Box t^{\sigma}_{\sigma}-2\hat\gamma^{\alpha\beta}\hat\nabla_{\sigma}t_{\rho\delta}\hat\nabla^{\delta}t^{\rho\sigma}+3\hat\gamma^{\alpha\beta}\hat\nabla_{\delta}t_{\rho\sigma}\hat\nabla^{\delta}t^{\rho\sigma}\notag\\
	&\quad+4\hat\nabla_{\rho}t^{\beta\rho}\hat\nabla^{\alpha}t^{\sigma}_{\sigma}+4t^{\rho}_{\rho}\hat\nabla^{\alpha}\hat\nabla_{\sigma}t^{\beta\sigma}-4t^{\rho\sigma}\hat\nabla^{\alpha}\hat\nabla^{\beta}t_{\rho\sigma}-8\hat\nabla^{\sigma}t^{\alpha\rho}\hat\nabla^{\beta}t_{\rho\sigma}\nonumber\\
	&\quad+4\hat\nabla^{\alpha}t^{\rho\sigma}\hat\nabla^{\beta}t_{\rho\sigma}+4\hat\nabla_{\rho}t^{\alpha\rho}\hat\nabla^{\beta}t^{\sigma}_{\sigma}-2\hat\nabla^{\alpha}t^{\rho}_{\rho}\hat\nabla^{\beta}t^{\sigma}_{\sigma}-2\hat\nabla_{\rho}t^{\sigma}_{\sigma}(\hat\nabla^{\rho}t^{\alpha\beta}-2\hat\nabla^{\beta}t^{\alpha\rho})\notag\\
	&\quad-8\hat\nabla_{\sigma}t_{\rho}^{\sigma}\hat\nabla^{\beta}t^{\alpha\rho}-4t^{\rho}_{\rho}\hat\nabla^{\beta}\hat\nabla_{\sigma}t^{\alpha\sigma}+2t^{\rho}_{\rho}\hat\nabla^{\beta}\hat\nabla^{\alpha}t^{\sigma}_{\sigma}\Big)\Big](y)+O(\lambda^3),\label{saddle-point metric up to second order}
\end{align}
where
\begin{align}
	t_{\mu\nu}(x)=\frac{1}{l^2}\int d^dx'\sqrt{\hat\gamma(x')}G_{\mu\nu\rho\sigma}(x,x')\Big[\hat T^{\rho\sigma}-\frac{1}{d-2}\hat T^{\alpha}_{\alpha}\hat\gamma^{\rho\sigma}\Big](x').
\end{align}
Plugging (\ref{saddle-point metric up to second order}) into the gravitational action, we obtain the second order contribution to the non-local stress-tensor deformed action induced by Einstein gravity:
\begin{align}
	&\quad S^{(\lambda)}_{[2]}[\hat\gamma,\psi]\notag\\
	&=-\frac{\lambda^2l^{2}}{2}\int d^dx\sqrt{\hat\gamma}\bigg[\frac{2}{l^2}t_{\mu\nu}t_{\rho\sigma}M^{\mu\nu\rho\sigma}+(t^{\rho}_{\rho})^2(\hat\nabla_{\nu}\hat\nabla_{\mu}t^{\mu\nu}-\hat\Box t^{\mu}_{\mu})-2t^{\mu\nu}(3\hat\nabla_{\mu}t^{\rho\sigma}\hat\nabla_{\nu}t_{\rho\sigma}\nonumber\\
	&\quad-\hat\nabla_{\mu}t^{\rho}_{\rho}\hat\nabla_{\nu}t^{\sigma}_{\sigma}+4\hat\nabla_{\nu}t^{\sigma}_{\sigma}\hat\nabla_{\rho}t_{\mu}^{\rho}+4\hat\nabla_{\nu}t_{\mu}^{\rho}\hat\nabla_{\rho}t^{\sigma}_{\sigma}+4t_{\mu}^{\rho}\hat\nabla_{\rho}\hat\nabla_{\nu}t^{\sigma}_{\sigma}-4t_{\mu}^{\rho}\hat\nabla_{\rho}\hat\nabla_{\sigma}t_{\nu}^{\sigma}-2\hat\nabla_{\rho}t^{\sigma}_{\sigma}\hat\nabla^{\rho}t_{\mu\nu}\notag\\
	&\quad-4\hat\nabla_{\rho}t_{\mu}^{\rho}\hat\nabla_{\sigma}t_{\nu}^{\sigma}-8\hat\nabla_{\nu}t_{\mu}^{\rho}\hat\nabla_{\sigma}t_{\rho}^{\sigma}+4\hat\nabla^{\rho}t_{\mu\nu}\hat\nabla_{\sigma}t_{\rho}^{\sigma}-4t^{\rho\sigma}\hat\nabla_{\sigma}\hat\nabla_{\nu}t_{\mu\rho}+4t^{\rho\sigma}\hat\nabla_{\sigma}\hat\nabla_{\rho}t_{\mu\nu}\nonumber\\
	&\quad-4t_{\mu}^{\rho}\hat\nabla_{\sigma}\hat\nabla_{\rho}t_{\nu}^{\sigma}+t_{\mu\nu}\hat\nabla_{\sigma}\hat\nabla_{\rho}t^{\rho\sigma}+4t_{\mu}^{\rho}\hat\Box t_{\nu\rho}-t_{\mu\nu}\hat\Box t^{\rho}_{\rho}-4\hat\nabla_{\nu}t_{\rho\sigma}\hat\nabla^{\sigma}t_{\mu}^{\rho}-2\hat\nabla_{\rho}t_{\nu\sigma}\hat\nabla^{\sigma}t_{\mu}^{\rho}\notag\\
	&\quad+6\hat\nabla_{\sigma}t_{\nu\rho}\hat\nabla^{\sigma}t_{\mu}^{\rho})-t^{\mu}_{\mu}(\hat\nabla_{\rho}t^{\sigma}_{\sigma}\hat\nabla^{\rho}t^{\nu}_{\nu}+4\hat\nabla_{\nu}t^{\nu\rho}\hat\nabla_{\sigma}t_{\rho}^{\sigma}-4\hat\nabla^{\rho}t^{\nu}_{\nu}\hat\nabla_{\sigma}t_{\rho}^{\sigma}-4t^{\nu\rho}(\hat\nabla_{\rho}\hat\nabla_{\nu}t^{\sigma}_{\sigma}\notag\\
	&\quad-\hat\nabla_{\rho}\hat\nabla_{\sigma}t_{\nu}^{\sigma}-\hat\nabla_{\sigma}\hat\nabla_{\rho}t_{\nu}^{\sigma}+\hat\Box t_{\nu\rho})+2\hat\nabla_{\rho}t_{\nu\sigma}\hat\nabla^{\sigma}t^{\nu\rho}-3\hat\nabla_{\sigma}t_{\nu\rho}\hat\nabla^{\sigma}t^{\nu\rho})\bigg].
\end{align}

\bibliographystyle{JHEP}
\bibliography{scalartoLiouville.bib}

@article{Adami:2025pqr,
  title={Gravity is induced by renormalization group flow},
  author={Adami, H and Sheikh-Jabbari, MM and Taghiloo, V},
  journal={arXiv preprint arXiv:2508.09633},
  year={2025}
}

@article{Weinberg:1980kq,
    author = "Weinberg, Steven and Witten, Edward",
    title = "{Limits on Massless Particles}",
    reportNumber = "HUTP-80/A056",
    doi = "10.1016/0370-2693(80)90212-9",
    journal = "Phys. Lett. B",
    volume = "96",
    pages = "59--62",
    year = "1980"
}

@article{Tsujikawa:2007xu,
    author = "Tsujikawa, Shinji",
    title = "{Observational signatures of $f(R)$ dark energy models that satisfy cosmological and local gravity constraints}",
    eprint = "0709.1391",
    archivePrefix = "arXiv",
    primaryClass = "astro-ph",
    doi = "10.1103/PhysRevD.77.023507",
    journal = "Phys. Rev. D",
    volume = "77",
    pages = "023507",
    year = "2008"
}

@article{DeFelice:2010aj,
    author = "De Felice, Antonio and Tsujikawa, Shinji",
    title = "{f(R) theories}",
    eprint = "1002.4928",
    archivePrefix = "arXiv",
    primaryClass = "gr-qc",
    doi = "10.12942/lrr-2010-3",
    journal = "Living Rev. Rel.",
    volume = "13",
    pages = "3",
    year = "2010"
}

@article{Bonelli:2018kik,
    author = "Bonelli, Giulio and Doroud, Nima and Zhu, Mengqi",
    title = "{$T \bar{T}$-deformations in closed form}",
    eprint = "1804.10967",
    archivePrefix = "arXiv",
    primaryClass = "hep-th",
    doi = "10.1007/JHEP06(2018)149",
    journal = "JHEP",
    volume = "06",
    pages = "149",
    year = "2018"
}

@article{Taylor:2018xcy,
    author = "Taylor, Marika",
    title = "{$T \bar{T}$ deformations in general dimensions}",
    eprint = "1805.10287",
    archivePrefix = "arXiv",
    primaryClass = "hep-th",
    doi = "10.4310/ATMP.2023.v27.n1.a2",
    journal = "Adv. Theor. Math. Phys.",
    volume = "27",
    number = "1",
    pages = "37--63",
    year = "2023"
}

@article{Conti:2022egv,
    author = "Conti, Riccardo and Romano, Jacopo and Tateo, Roberto",
    title = "{Metric approach to a $ \mathrm{T}\overline{\mathrm{T}} $-like deformation in arbitrary dimensions}",
    eprint = "2206.03415",
    archivePrefix = "arXiv",
    primaryClass = "hep-th",
    doi = "10.1007/JHEP09(2022)085",
    journal = "JHEP",
    volume = "09",
    pages = "085",
    year = "2022"
}

@article{Visser:2002ew,
    author = "Visser, Matt",
    editor = "Ahluwalia, Dharam Vir and Dadhich, N. K.",
    title = "{Sakharov's induced gravity: A Modern perspective}",
    eprint = "gr-qc/0204062",
    archivePrefix = "arXiv",
    doi = "10.1142/S0217732302006886",
    journal = "Mod. Phys. Lett. A",
    volume = "17",
    pages = "977--992",
    year = "2002"
}

@article{Maldacena:1997re,
    author = "Maldacena, Juan Martin",
    title = "{The Large $N$ limit of superconformal field theories and supergravity}",
    eprint = "hep-th/9711200",
    archivePrefix = "arXiv",
    reportNumber = "HUTP-97-A097, HUTP-98-A097",
    doi = "10.4310/ATMP.1998.v2.n2.a1",
    journal = "Adv. Theor. Math. Phys.",
    volume = "2",
    pages = "231--252",
    year = "1998"
}

@article{Verlinde:2010hp,
    author = "Verlinde, Erik P.",
    title = "{On the Origin of Gravity and the Laws of Newton}",
    eprint = "1001.0785",
    archivePrefix = "arXiv",
    primaryClass = "hep-th",
    doi = "10.1007/JHEP04(2011)029",
    journal = "JHEP",
    volume = "04",
    pages = "029",
    year = "2011"
}

@article{Witten:1998qj,
    author = "Witten, Edward",
    title = "{Anti de Sitter space and holography}",
    eprint = "hep-th/9802150",
    archivePrefix = "arXiv",
    reportNumber = "IASSNS-HEP-98-15",
    doi = "10.4310/ATMP.1998.v2.n2.a2",
    journal = "Adv. Theor. Math. Phys.",
    volume = "2",
    pages = "253--291",
    year = "1998"
}

@article{Betzios:2020sro,
    author = "Betzios, Panos and Kiritsis, Elias and Niarchos, Vasilis",
    title = "{Emergent gravity from hidden sectors and TT deformations}",
    eprint = "2010.04729",
    archivePrefix = "arXiv",
    primaryClass = "hep-th",
    reportNumber = "CCTP-2020-11, ITCP-2020/11, DCPT-20/15",
    doi = "10.1007/JHEP02(2021)202",
    journal = "JHEP",
    volume = "02",
    pages = "202",
    year = "2021"
}

@article{Dubovsky:2017cnj,
    author = "Dubovsky, Sergei and Gorbenko, Victor and Mirbabayi, Mehrdad",
    title = "{Asymptotic fragility, near AdS$_{2}$ holography and $ T\overline{T} $}",
    eprint = "1706.06604",
    archivePrefix = "arXiv",
    primaryClass = "hep-th",
    doi = "10.1007/JHEP09(2017)136",
    journal = "JHEP",
    volume = "09",
    pages = "136",
    year = "2017"
}

@article{Dubovsky:2018bmo,
    author = "Dubovsky, Sergei and Gorbenko, Victor and Hern{\'a}ndez-Chifflet, Guzm{\'a}n",
    title = "{$ T\overline{T} $ partition function from topological gravity}",
    eprint = "1805.07386",
    archivePrefix = "arXiv",
    primaryClass = "hep-th",
    doi = "10.1007/JHEP09(2018)158",
    journal = "JHEP",
    volume = "09",
    pages = "158",
    year = "2018"
}

@article{Tolley:2019nmm,
    author = "Tolley, Andrew J.",
    title = "{$ T\overline{T} $ deformations, massive gravity and non-critical strings}",
    eprint = "1911.06142",
    archivePrefix = "arXiv",
    primaryClass = "hep-th",
    reportNumber = "Imperial/TP/2019/AT/01",
    doi = "10.1007/JHEP06(2020)050",
    journal = "JHEP",
    volume = "06",
    pages = "050",
    year = "2020"
}

@article{He:2025ppz,
    author = "He, Song and Li, Yi and Ouyang, Hao and Sun, Yuan",
    title = "{$T\overline{T}$ deformation: Introduction and some recent advances}",
    eprint = "2503.09997",
    archivePrefix = "arXiv",
    primaryClass = "hep-th",
    doi = "10.1007/s11433-025-2708-2",
    journal = "Sci. China Phys. Mech. Astron.",
    volume = "68",
    number = "10",
    pages = "101001",
    year = "2025"
}

@article{Cavaglia:2016oda,
	author = "Cavagli\`a, Andrea and Negro, Stefano and Sz\'ecs\'enyi, Istv\'an M. and Tateo, Roberto",
	title = "{$T \bar{T}$-deformed 2D Quantum Field Theories}",
	eprint = "1608.05534",
	archivePrefix = "arXiv",
	primaryClass = "hep-th",
	doi = "10.1007/JHEP10(2016)112",
	journal = "JHEP",
	volume = "10",
	pages = "112",
	year = "2016"
}

@article{Babaei-Aghbolagh:2024hti,
	author = "Babaei-Aghbolagh, H. and He, Song and Morone, Tommaso and Ouyang, Hao and Tateo, Roberto",
	title = "{Geometric Formulation of Generalized Root-TT\textasciimacron{} Deformations}",
	eprint = "2405.03465",
	archivePrefix = "arXiv",
	primaryClass = "hep-th",
	doi = "10.1103/PhysRevLett.133.111602",
	journal = "Phys. Rev. Lett.",
	volume = "133",
	number = "11",
	pages = "111602",
	year = "2024"
}

@article{Jacobson:1995ab,
    author = "Jacobson, Ted",
    title = "{Thermodynamics of space-time: The Einstein equation of state}",
    eprint = "gr-qc/9504004",
    archivePrefix = "arXiv",
    reportNumber = "UMDGR-95-114",
    doi = "10.1103/PhysRevLett.75.1260",
    journal = "Phys. Rev. Lett.",
    volume = "75",
    pages = "1260--1263",
    year = "1995"
}

@article{Padmanabhan:2010xh,
    author = "Padmanabhan, T.",
    title = "{Surface Density of Spacetime Degrees of Freedom from Equipartition Law in theories of Gravity}",
    eprint = "1003.5665",
    archivePrefix = "arXiv",
    primaryClass = "gr-qc",
    doi = "10.1103/PhysRevD.81.124040",
    journal = "Phys. Rev. D",
    volume = "81",
    pages = "124040",
    year = "2010"
}

@article{Vassilevich:2003xt,
	author = "Vassilevich, D. V.",
	title = "{Heat kernel expansion: User's manual}",
	eprint = "hep-th/0306138",
	archivePrefix = "arXiv",
	doi = "10.1016/j.physrep.2003.09.002",
	journal = "Phys. Rept.",
	volume = "388",
	pages = "279--360",
	year = "2003"
}

@article{Kawamoto:2025oko,
    author = "Kawamoto, Taishi and Maeda, Ryota and Nakamura, Nanami and Takayanagi, Tadashi",
    title = "{Traversable AdS wormhole via non-local double trace or Janus deformation}",
    eprint = "2502.03531",
    archivePrefix = "arXiv",
    primaryClass = "hep-th",
    reportNumber = "YITP-25-09",
    doi = "10.1007/JHEP04(2025)086",
    journal = "JHEP",
    volume = "04",
    pages = "086",
    year = "2025"
}

@article{Giombi:2008vd,
    author = "Giombi, Simone and Maloney, Alexander and Yin, Xi",
    title = "{One-loop Partition Functions of 3D Gravity}",
    eprint = "0804.1773",
    archivePrefix = "arXiv",
    primaryClass = "hep-th",
    doi = "10.1088/1126-6708/2008/08/007",
    journal = "JHEP",
    volume = "08",
    pages = "007",
    year = "2008"
}

@article{Barvinsky:2021ijq,
    author = "Barvinsky, Andrei O. and Wachowski, Wladyslaw",
    title = "{Heat kernel expansion for higher order minimal and nonminimal operators}",
    eprint = "2112.03062",
    archivePrefix = "arXiv",
    primaryClass = "hep-th",
    doi = "10.1103/PhysRevD.105.065013",
    journal = "Phys. Rev. D",
    volume = "105",
    number = "6",
    pages = "065013",
    year = "2022",
    note = "[Erratum: Phys.Rev.D 110, 089901 (2024)]"
}

@article{Morone:2024ffm,
	author = "Morone, Tommaso and Negro, Stefano and Tateo, Roberto",
	title = "{Gravity and TT‾ flows in higher dimensions}",
	eprint = "2401.16400",
	archivePrefix = "arXiv",
	primaryClass = "hep-th",
	doi = "10.1016/j.nuclphysb.2024.116605",
	journal = "Nucl. Phys. B",
	volume = "1005",
	pages = "116605",
	year = "2024"
}

@article{Brizio:2024arr,
	author = "Brizio, Nicol\`o and Morone, Tommaso and Tateo, Roberto",
	title = "{Stress-energy tensor deformations, Ricci flows and black holes}",
	eprint = "2408.06031",
	archivePrefix = "arXiv",
	primaryClass = "hep-th",
	month = "8",
	year = "2024"
}

@article{Morone:2024sdg,
	author = "Morone, Tommaso and Tateo, Roberto",
	title = "{Solutions to the Ricci Flow via Einstein Field Equations}",
	eprint = "2411.10265",
	archivePrefix = "arXiv",
	primaryClass = "hep-th",
	month = "11",
	year = "2024"
}

@article{Gullu:2010em,
    author = "Gullu, Ibrahim and Sisman, Tahsin Cagri and Tekin, Bayram",
    title = "{Unitarity analysis of general Born-Infeld gravity theories}",
    eprint = "1010.2411",
    archivePrefix = "arXiv",
    primaryClass = "hep-th",
    doi = "10.1103/PhysRevD.82.124023",
    journal = "Phys. Rev. D",
    volume = "82",
    pages = "124023",
    year = "2010"
}

@article{Altas:2019qcv,
    author = "Altas, Emel and Tekin, Bayram",
    title = "{Second Order Perturbation Theory in General Relativity: Taub Charges as Integral Constraints}",
    eprint = "1903.11982",
    archivePrefix = "arXiv",
    primaryClass = "hep-th",
    doi = "10.1103/PhysRevD.99.104078",
    journal = "Phys. Rev. D",
    volume = "99",
    number = "10",
    pages = "104078",
    year = "2019"
}

@article{Christensen:1979iy,
    author = "Christensen, S. M. and Duff, M. J.",
    title = "{Quantizing Gravity with a Cosmological Constant}",
    reportNumber = "NSF-ITP-79-01",
    doi = "10.1016/0550-3213(80)90423-X",
    journal = "Nucl. Phys. B",
    volume = "170",
    pages = "480--506",
    year = "1980"
}

@article{Ran:2024vgl,
	author = "Ran, Xi-Yang and Hao, Feng and Yamada, Masatoshi",
	title = "{Geometric realization via irrelevant deformations induced by the stress-energy tensor}",
	eprint = "2410.02537",
	archivePrefix = "arXiv",
	primaryClass = "hep-th",
	doi = "10.1103/PhysRevD.111.085033",
	journal = "Phys. Rev. D",
	volume = "111",
	number = "8",
	pages = "085033",
	year = "2025"
}

@article{Sotiriou:2008rp,
    author = "Sotiriou, Thomas P. and Faraoni, Valerio",
    title = "{f(R) Theories Of Gravity}",
    eprint = "0805.1726",
    archivePrefix = "arXiv",
    primaryClass = "gr-qc",
    doi = "10.1103/RevModPhys.82.451",
    journal = "Rev. Mod. Phys.",
    volume = "82",
    pages = "451--497",
    year = "2010"
}

@article{Cardy:2018sdv,
    author = "Cardy, John",
    title = "{The $ T\overline{T} $ deformation of quantum field theory as random geometry}",
    eprint = "1801.06895",
    archivePrefix = "arXiv",
    primaryClass = "hep-th",
    doi = "10.1007/JHEP10(2018)186",
    journal = "JHEP",
    volume = "10",
    pages = "186",
    year = "2018"
}

@book{Polchinski:1998rq,
    author = "Polchinski, J.",
    title = "{String theory. Vol. 1: An introduction to the bosonic string}",
    doi = "10.1017/CBO9780511816079",
    isbn = "978-0-511-25227-3, 978-0-521-67227-6, 978-0-521-63303-1",
    publisher = "Cambridge University Press",
    series = "Cambridge Monographs on Mathematical Physics",
    month = "12",
    year = "2007"
}

@article{Ferko:2024zth,
    author = "Ferko, Christian and Kuzenko, Sergei M. and Lechner, Kurt and Sorokin, Dmitri P. and Tartaglino-Mazzucchelli, Gabriele",
    title = "{Interacting chiral form field theories and $ T\overline{T} $-like flows in six and higher dimensions}",
    eprint = "2402.06947",
    archivePrefix = "arXiv",
    primaryClass = "hep-th",
    doi = "10.1007/JHEP05(2024)320",
    journal = "JHEP",
    volume = "05",
    pages = "320",
    year = "2024"
}

@techreport{Zamolodchikov:2004ce,
  title={Expectation value of composite field T anti-T in two-dimensional quantum field theory},
  author={Zamolodchikov, Alexander B},
  year={2004}
}

@article{Smirnov:2016lqw,
    author = "Smirnov, F. A. and Zamolodchikov, A. B.",
    title = "{On space of integrable quantum field theories}",
    eprint = "1608.05499",
    archivePrefix = "arXiv",
    primaryClass = "hep-th",
    doi = "10.1016/j.nuclphysb.2016.12.014",
    journal = "Nucl. Phys. B",
    volume = "915",
    pages = "363--383",
    year = "2017"
}

@article{Li:2025lpa,
	author = "Li, Yun-Ze and Xie, Yunfei and He, Song",
	title = "{Emergent classical gravity as stress-tensor deformed field theories}",
	eprint = "2508.15461",
	archivePrefix = "arXiv",
	primaryClass = "hep-th",
	month = "8",
	year = "2025"
}

@article{Babaei-Aghbolagh:2022leo,
	author = "Babaei-Aghbolagh, H. and Babaei Velni, Komeil and Mahdavian Yekta, Davood and Mohammadzadeh, Hosein",
	title = "{Marginal TT{\textasciimacron}-like deformation and modified Maxwell theories in two dimensions}",
	eprint = "2206.12677",
	archivePrefix = "arXiv",
	primaryClass = "hep-th",
	doi = "10.1103/PhysRevD.106.086022",
	journal = "Phys. Rev. D",
	volume = "106",
	number = "8",
	pages = "086022",
	year = "2022"
}

@article{DeWitt:1964mxt,
    author = "DeWitt, Bryce S.",
    editor = "DeWitt, C. and DeWitt, B.",
    title = "{Dynamical theory of groups and fields}",
    journal = "Conf. Proc. C",
    volume = "630701",
    pages = "585--820",
    year = "1964"
}

@article{Barvinsky:1985an,
    author = "Barvinsky, A. O. and Vilkovisky, G. A.",
    title = "{The Generalized Schwinger-Dewitt Technique in Gauge Theories and Quantum Gravity}",
    doi = "10.1016/0370-1573(85)90148-6",
    journal = "Phys. Rept.",
    volume = "119",
    pages = "1--74",
    year = "1985"
}

@article{Seeley:1967ea,
    author = "Seeley, R. T.",
    title = "{Complex powers of an elliptic operator}",
    journal = "Proc. Symp. Pure Math.",
    volume = "10",
    pages = "288--307",
    year = "1967"
}

@book{berline2003heat,
  title={Heat Kernels and Dirac Operators},
  author={Berline, N. and Getzler, E. and Vergne, M.},
  isbn={9783540200628},
  lccn={91039973},
  series={Grundlehren Text Editions},
  url={https://books.google.com.hk/books?id=_e2FjvLbO94C},
  year={2003},
  publisher={Springer Berlin Heidelberg}
}

@article{Hartman:2023ccw,
    author = "Hartman, Thomas and Mathys, Gr{\'e}goire",
    title = "{Null energy constraints on two-dimensional RG flows}",
    eprint = "2310.15217",
    archivePrefix = "arXiv",
    primaryClass = "hep-th",
    doi = "10.1007/JHEP01(2024)102",
    journal = "JHEP",
    volume = "01",
    pages = "102",
    year = "2024"
}

@article{Hartman:2023qdn,
    author = "Hartman, Thomas and Mathys, Gr{\'e}goire",
    title = "{Averaged null energy and the renormalization group}",
    eprint = "2309.14409",
    archivePrefix = "arXiv",
    primaryClass = "hep-th",
    doi = "10.1007/JHEP12(2023)139",
    journal = "JHEP",
    volume = "12",
    pages = "139",
    year = "2023"
}

@article{Borsato:2022tmu,
    author = "Borsato, Riccardo and Ferko, Christian and Sfondrini, Alessandro",
    title = "{Classical integrability of root-TT{\textasciimacron} flows}",
    eprint = "2209.14274",
    archivePrefix = "arXiv",
    primaryClass = "hep-th",
    doi = "10.1103/PhysRevD.107.086011",
    journal = "Phys. Rev. D",
    volume = "107",
    number = "8",
    pages = "086011",
    year = "2023"
}

@article{Babaei-Aghbolagh:2022itg,
    author = "Babaei-Aghbolagh, H. and Babaei Velni, Komeil and Yekta, Davood Mahdavian and Mohammadzadeh, H.",
    title = "{Manifestly SL(2, R) Duality-Symmetric Forms in ModMax Theory}",
    eprint = "2210.13196",
    archivePrefix = "arXiv",
    primaryClass = "hep-th",
    doi = "10.1007/JHEP12(2022)147",
    journal = "JHEP",
    volume = "12",
    pages = "147",
    year = "2022"
}

@article{Ferko:2023ozb,
    author = "Ferko, Christian and Gupta, Alisha",
    title = "{ModMax oscillators and root-TT{\textasciimacron}-like flows in supersymmetric quantum mechanics}",
    eprint = "2306.14575",
    archivePrefix = "arXiv",
    primaryClass = "hep-th",
    doi = "10.1103/PhysRevD.108.046013",
    journal = "Phys. Rev. D",
    volume = "108",
    number = "4",
    pages = "046013",
    year = "2023"
}

@article{Conti:2018jho,
    author = "Conti, Riccardo and Iannella, Leonardo and Negro, Stefano and Tateo, Roberto",
    title = "{Generalised Born-Infeld models, Lax operators and the $ \mathrm{T}\overline{\mathrm{T}} $ perturbation}",
    eprint = "1806.11515",
    archivePrefix = "arXiv",
    primaryClass = "hep-th",
    doi = "10.1007/JHEP11(2018)007",
    journal = "JHEP",
    volume = "11",
    pages = "007",
    year = "2018"
}

@article{Ferko:2022iru,
    author = "Ferko, Christian and Smith, Liam and Tartaglino-Mazzucchelli, Gabriele",
    title = "{On Current-Squared Flows and ModMax Theories}",
    eprint = "2203.01085",
    archivePrefix = "arXiv",
    primaryClass = "hep-th",
    doi = "10.21468/SciPostPhys.13.2.012",
    journal = "SciPost Phys.",
    volume = "13",
    number = "2",
    pages = "012",
    year = "2022"
}

@article{Ferko:2023wyi,
    author = "Ferko, Christian and Kuzenko, Sergei M. and Smith, Liam and Tartaglino-Mazzucchelli, Gabriele",
    title = "{Duality-invariant nonlinear electrodynamics and stress tensor flows}",
    eprint = "2309.04253",
    archivePrefix = "arXiv",
    primaryClass = "hep-th",
    doi = "10.1103/PhysRevD.108.106021",
    journal = "Phys. Rev. D",
    volume = "108",
    number = "10",
    pages = "106021",
    year = "2023"
}

@article{Ferko:2023ruw,
    author = "Ferko, Christian and Smith, Liam and Tartaglino-Mazzucchelli, Gabriele",
    title = "{Stress Tensor flows, birefringence in non-linear electrodynamics and supersymmetry}",
    eprint = "2301.10411",
    archivePrefix = "arXiv",
    primaryClass = "hep-th",
    doi = "10.21468/SciPostPhys.15.5.198",
    journal = "SciPost Phys.",
    volume = "15",
    number = "5",
    pages = "198",
    year = "2023"
}

@article{Avramidi:2001ns,
    author = "Avramidi, Ivan G.",
    editor = "Esposito, Giampiero and Miele, Gennaro and Preziosi, Bruno",
    title = "{Heat kernel approach in quantum field theory}",
    eprint = "math-ph/0107018",
    archivePrefix = "arXiv",
    doi = "10.1016/S0920-5632(01)01593-6",
    journal = "Nucl. Phys. B Proc. Suppl.",
    volume = "104",
    pages = "3--32",
    year = "2002"
}

@book{Avramidi:2000bm,
    author = "Avramidi, I. G.",
    title = "{Heat kernel and quantum gravity}",
    doi = "10.1007/3-540-46523-5",
    isbn = "978-3-540-67155-8",
    publisher = "Springer",
    address = "New York",
    volume = "64",
    year = "2000"
}

@article{Babaei-Aghbolagh:2025lko,
	author = "Babaei-Aghbolagh, H. and He, Song and Ouyang, Hao",
	title = "{Generalized TT{\textasciimacron}-like flows for scalar theories in two dimensions}",
	eprint = "2501.14583",
	archivePrefix = "arXiv",
	primaryClass = "hep-th",
	doi = "10.1103/58fz-mkxx",
	journal = "Phys. Rev. D",
	volume = "112",
	number = "6",
	pages = "066005",
	year = "2025"
}

@article{Callebaut:2019omt,
	author = "Callebaut, Nele and Kruthoff, Jorrit and Verlinde, Herman",
	title = "{$ T\overline{T} $ deformed CFT as a non-critical string}",
	eprint = "1910.13578",
	archivePrefix = "arXiv",
	primaryClass = "hep-th",
	doi = "10.1007/JHEP04(2020)084",
	journal = "JHEP",
	volume = "04",
	pages = "084",
	year = "2020"
}

@article{McGough:2016lol,
	author = "McGough, Lauren and Mezei, M{\'a}rk and Verlinde, Herman",
	title = "{Moving the CFT into the bulk with $ T\overline{T} $}",
	eprint = "1611.03470",
	archivePrefix = "arXiv",
	primaryClass = "hep-th",
	doi = "10.1007/JHEP04(2018)010",
	journal = "JHEP",
	volume = "04",
	pages = "010",
	year = "2018"
}

@article{Guica:2019nzm,
	author = "Guica, Monica and Monten, Ruben",
	title = "{$T\bar T$ and the mirage of a bulk cutoff}",
	eprint = "1906.11251",
	archivePrefix = "arXiv",
	primaryClass = "hep-th",
	doi = "10.21468/SciPostPhys.10.2.024",
	journal = "SciPost Phys.",
	volume = "10",
	number = "2",
	pages = "024",
	year = "2021"
}

@article{Giveon:2017nie,
	author = "Giveon, Amit and Itzhaki, Nissan and Kutasov, David",
	title = "{$ \mathrm{T}\overline{\mathrm{T}} $ and LST}",
	eprint = "1701.05576",
	archivePrefix = "arXiv",
	primaryClass = "hep-th",
	doi = "10.1007/JHEP07(2017)122",
	journal = "JHEP",
	volume = "07",
	pages = "122",
	year = "2017"
}

@article{Giveon:2017myj,
	author = "Giveon, Amit and Itzhaki, Nissan and Kutasov, David",
	title = "{A solvable irrelevant deformation of AdS$_{3}$/CFT$_{2}$}",
	eprint = "1707.05800",
	archivePrefix = "arXiv",
	primaryClass = "hep-th",
	doi = "10.1007/JHEP12(2017)155",
	journal = "JHEP",
	volume = "12",
	pages = "155",
	year = "2017"
}

@article{Chakraborty:2019mdf,
	author = "Chakraborty, Soumangsu and Giveon, Amit and Kutasov, David",
	title = "{$T\bar{T}$, $J\bar{T}$, $T\bar{J}$ and String Theory}",
	eprint = "1905.00051",
	archivePrefix = "arXiv",
	primaryClass = "hep-th",
	doi = "10.1088/1751-8121/ab3710",
	journal = "J. Phys. A",
	volume = "52",
	number = "38",
	pages = "384003",
	year = "2019"
}

@article{Apolo:2019zai,
	author = "Apolo, Luis and Detournay, Stephane and Song, Wei",
	title = "{TsT, $T\bar{T}$ and black strings}",
	eprint = "1911.12359",
	archivePrefix = "arXiv",
	primaryClass = "hep-th",
	doi = "10.1007/JHEP06(2020)109",
	journal = "JHEP",
	volume = "06",
	pages = "109",
	year = "2020"
}

@article{Baggio:2018gct,
	author = "Baggio, Marco and Sfondrini, Alessandro",
	title = "{Strings on NS-NS Backgrounds as Integrable Deformations}",
	eprint = "1804.01998",
	archivePrefix = "arXiv",
	primaryClass = "hep-th",
	doi = "10.1103/PhysRevD.98.021902",
	journal = "Phys. Rev. D",
	volume = "98",
	number = "2",
	pages = "021902",
	year = "2018"
}

@article{Dei:2018jyj,
	author = "Dei, Andrea and Sfondrini, Alessandro",
	title = "{Integrable S matrix, mirror TBA and spectrum for the stringy AdS$_{3}$ {\texttimes} S$^{3}$ {\texttimes} S$^{3}$ {\texttimes} S$^{1}$ WZW model}",
	eprint = "1812.08195",
	archivePrefix = "arXiv",
	primaryClass = "hep-th",
	doi = "10.1007/JHEP02(2019)072",
	journal = "JHEP",
	volume = "02",
	pages = "072",
	year = "2019"
}

@article{Ferko:2024ali,
	author = "Ferko, Christian and Smith, Liam",
	title = "{Infinite Family of Integrable Sigma Models Using Auxiliary Fields}",
	eprint = "2405.05899",
	archivePrefix = "arXiv",
	primaryClass = "hep-th",
	doi = "10.1103/PhysRevLett.133.131602",
	journal = "Phys. Rev. Lett.",
	volume = "133",
	number = "13",
	pages = "131602",
	year = "2024"
}

@article{Pozsgay:2019ekd,
	author = "Pozsgay, Bal{\'a}zs and Jiang, Yunfeng and Tak{\'a}cs, G{\'a}bor",
	title = "{$T\bar T$-deformation and long range spin chains}",
	eprint = "1911.11118",
	archivePrefix = "arXiv",
	primaryClass = "hep-th",
	reportNumber = "CERN-TH-2019-200",
	doi = "10.1007/JHEP03(2020)092",
	journal = "JHEP",
	volume = "03",
	pages = "092",
	year = "2020"
}

@article{Marchetto:2019yyt,
	author = "Marchetto, Enrico and Sfondrini, Alessandro and Yang, Zhou",
	title = "{$T\bar{T}$ Deformations and Integrable Spin Chains}",
	eprint = "1911.12315",
	archivePrefix = "arXiv",
	primaryClass = "hep-th",
	doi = "10.1103/PhysRevLett.124.100601",
	journal = "Phys. Rev. Lett.",
	volume = "124",
	number = "10",
	pages = "100601",
	year = "2020"
}

@article{Jiang:2020nnb,
	author = "Jiang, Yunfeng",
	title = "{$\mathrm{T}\overline{\mathrm{T}}$-deformed 1d Bose gas}",
	eprint = "2011.00637",
	archivePrefix = "arXiv",
	primaryClass = "hep-th",
	reportNumber = "CERN-TH-2020-183",
	doi = "10.21468/SciPostPhys.12.6.191",
	journal = "SciPost Phys.",
	volume = "12",
	number = "6",
	pages = "191",
	year = "2022"
}

@article{Cardy:2020olv,
	author = "Cardy, John and Doyon, Benjamin",
	title = "{$ T\overline{T} $ deformations and the width of fundamental particles}",
	eprint = "2010.15733",
	archivePrefix = "arXiv",
	primaryClass = "hep-th",
	doi = "10.1007/JHEP04(2022)136",
	journal = "JHEP",
	volume = "04",
	pages = "136",
	year = "2022"
}

@article{Iliesiu:2020zld,
	author = "Iliesiu, Luca V. and Kruthoff, Jorrit and Turiaci, Gustavo J. and Verlinde, Herman",
	title = "{JT gravity at finite cutoff}",
	eprint = "2004.07242",
	archivePrefix = "arXiv",
	primaryClass = "hep-th",
	doi = "10.21468/SciPostPhys.9.2.023",
	journal = "SciPost Phys.",
	volume = "9",
	pages = "023",
	year = "2020"
}

@article{Okumura:2020dzb,
	author = "Okumura, Suguru and Yoshida, Kentaroh",
	title = "{$T\bar{T}$-deformation and Liouville gravity}",
	eprint = "2003.14148",
	archivePrefix = "arXiv",
	primaryClass = "hep-th",
	reportNumber = "KUNS-2810",
	doi = "10.1016/j.nuclphysb.2020.115083",
	journal = "Nucl. Phys. B",
	volume = "957",
	pages = "115083",
	year = "2020"
}

@article{Ebert:2022ehb,
	author = "Ebert, Stephen and Ferko, Christian and Sun, Hao-Yu and Sun, Zhengdi",
	title = "{$T\bar{T}$ in JT Gravity and BF Gauge Theory}",
	eprint = "2205.07817",
	archivePrefix = "arXiv",
	primaryClass = "hep-th",
	doi = "10.21468/SciPostPhys.13.4.096",
	journal = "SciPost Phys.",
	volume = "13",
	number = "4",
	pages = "096",
	year = "2022"
}

@article{Bhattacharyya:2023gvg,
	author = "Bhattacharyya, Arpan and Ghosh, Saptaswa and Pal, Sounak",
	title = "{Aspects of TT{\textasciimacron}+JT{\textasciimacron} deformed Schwarzian: From gravity partition function to late-time spectral form factor}",
	eprint = "2309.16658",
	archivePrefix = "arXiv",
	primaryClass = "hep-th",
	doi = "10.1103/PhysRevD.110.126015",
	journal = "Phys. Rev. D",
	volume = "110",
	number = "12",
	pages = "126015",
	year = "2024"
}

@article{Ferko:2022cix,
	author = "Ferko, Christian and Sfondrini, Alessandro and Smith, Liam and Tartaglino-Mazzucchelli, Gabriele",
	title = "{Root-$T \bar T$ Deformations in Two-Dimensional Quantum Field Theories}",
	eprint = "2206.10515",
	archivePrefix = "arXiv",
	primaryClass = "hep-th",
	doi = "10.1103/PhysRevLett.129.201604",
	journal = "Phys. Rev. Lett.",
	volume = "129",
	number = "20",
	pages = "201604",
	year = "2022"
}

@article{Babaei-Aghbolagh:2022uij,
	author = "Babaei-Aghbolagh, H. and Velni, Komeil Babaei and Yekta, Davood Mahdavian and Mohammadzadeh, H.",
	title = "{Emergence of non-linear electrodynamic theories from TT{\textasciimacron}-like deformations}",
	eprint = "2202.11156",
	archivePrefix = "arXiv",
	primaryClass = "hep-th",
	reportNumber = "IPM/P-2022/13",
	doi = "10.1016/j.physletb.2022.137079",
	journal = "Phys. Lett. B",
	volume = "829",
	pages = "137079",
	year = "2022"
}

@article{Babaei-Aghbolagh:2024uqp,
	author = "Babaei-Aghbolagh, H. and He, Song and Ouyang, Hao",
	title = "{Generalized $ T\overline{T} $-like deformations in duality-invariant nonlinear electrodynamic theories}",
	eprint = "2407.03698",
	archivePrefix = "arXiv",
	primaryClass = "hep-th",
	doi = "10.1007/JHEP09(2024)137",
	journal = "JHEP",
	volume = "09",
	pages = "137",
	year = "2024"
}

@article{Ebert:2023tih,
	author = "Ebert, Stephen and Ferko, Christian and Sun, Zhengdi",
	title = "{Root-TT{\textasciimacron} deformed boundary conditions in holography}",
	eprint = "2304.08723",
	archivePrefix = "arXiv",
	primaryClass = "hep-th",
	doi = "10.1103/PhysRevD.107.126022",
	journal = "Phys. Rev. D",
	volume = "107",
	number = "12",
	pages = "126022",
	year = "2023"
}

@article{Conti:2018tca,
	author = "Conti, Riccardo and Negro, Stefano and Tateo, Roberto",
	title = "{The $ \mathrm{T}\overline{\mathrm{T}} $ perturbation and its geometric interpretation}",
	eprint = "1809.09593",
	archivePrefix = "arXiv",
	primaryClass = "hep-th",
	doi = "10.1007/JHEP02(2019)085",
	journal = "JHEP",
	volume = "02",
	pages = "085",
	year = "2019"
}

@article{Caputa:2020lpa,
	author = "Caputa, Pawel and Caputa, Pawel and Datta, Shouvik and Datta, Shouvik and Jiang, Yunfeng and Jiang, Yunfeng and Kraus, Per and Kraus, Per",
	title = "{Geometrizing $ T\overline{T} $}",
	eprint = "2011.04664",
	archivePrefix = "arXiv",
	primaryClass = "hep-th",
	reportNumber = "CERN-TH-2020-188",
	doi = "10.1007/JHEP03(2021)140",
	journal = "JHEP",
	volume = "03",
	pages = "140",
	year = "2021",
	note = "[Erratum: JHEP 09, 110 (2022)]"
}

@article{hamilton1982three,
	title={Three-manifolds with positive Ricci curvature},
	author={Hamilton, Richard S},
	journal={Journal of Differential geometry},
	volume={17},
	number={2},
	pages={255--306},
	year={1982},
	publisher={Lehigh University}
}

@article{catino2017ricci,
	title={The Ricci--Bourguignon flow},
	author={Catino, Giovanni and Cremaschi, Laura and Djadli, Zindine and Mantegazza, Carlo and Mazzieri, Lorenzo},
	journal={Pacific Journal of Mathematics},
	volume={287},
	number={2},
	pages={337--370},
	year={2017},
	publisher={Mathematical Sciences Publishers}
}

@article{Tsolakidis:2024wut,
	author = "Tsolakidis, Evangelos",
	title = "{Massive gravity generalization of $ T\overline{T} $ deformations}",
	eprint = "2405.07967",
	archivePrefix = "arXiv",
	primaryClass = "hep-th",
	doi = "10.1007/JHEP09(2024)167",
	journal = "JHEP",
	volume = "09",
	pages = "167",
	year = "2024"
}

@article{Bandos:2020jsw,
	author = "Bandos, Igor and Lechner, Kurt and Sorokin, Dmitri and Townsend, Paul K.",
	title = "{A non-linear duality-invariant conformal extension of Maxwell's equations}",
	eprint = "2007.09092",
	archivePrefix = "arXiv",
	primaryClass = "hep-th",
	doi = "10.1103/PhysRevD.102.121703",
	journal = "Phys. Rev. D",
	volume = "102",
	pages = "121703",
	year = "2020"
}

@article{born1934foundations,
	title={Foundations of the new field theory},
	author={Born, Max and Infeld, Leopold},
	journal={Proceedings of the Royal Society of London. Series A, Containing Papers of a Mathematical and Physical Character},
	volume={144},
	number={852},
	pages={425--451},
	year={1934},
	publisher={The Royal Society London}
}

@article{barbashov1966scattering,
	title={Scattering of two plane electromagnetic waves in the non-linear Born-Infeld electrodynamics},
	author={Barbashov, BM and Chernikov, NA},
	journal={Communications in Mathematical Physics},
	volume={3},
	number={5},
	pages={313--322},
	year={1966},
	publisher={Springer}
}

@article{Babaei-Aghbolagh:2025uoz,
	author = "Babaei-Aghbolagh, H. and Chen, Bin and He, Song",
	title = "{Root-TT{\textasciimacron} flows unify 4D duality-invariant electrodynamics and 2D integrable sigma models}",
	eprint = "2507.22808",
	archivePrefix = "arXiv",
	primaryClass = "hep-th",
	doi = "10.1103/1r4p-3r5q",
	journal = "Phys. Rev. D",
	volume = "112",
	number = "10",
	pages = "L101702",
	year = "2025"
}

@article{Nix:2025plr,
    author = "Nix, Alexia and Tsolakidis, Evangelos",
    title = "{Massive gravity applications for $T\overline{T}$ deformations}",
    eprint = "2512.23533",
    archivePrefix = "arXiv",
    primaryClass = "hep-th",
    month = "12",
    year = "2025"
}

@article{Brizio:2026ynw,
    author = "Brizio, Nicol{\`o} and Kade, Moritz and Sfondrini, Alessandro and Sorokin, Dmitri P.",
    title = "{More on $T \overline{T}$-like deformations in higher dimensions}",
    eprint = "2602.24058",
    archivePrefix = "arXiv",
    primaryClass = "hep-th",
    month = "2",
    year = "2026"
}

@article{Sakharov:1967nyk,
	author = "Sakharov, A. D.",
	title = "{Vacuum Quantum Fluctuations in Curved Space and the Theory of Gravitation}",
	doi = "10.1023/A:1001947813563",
	journal = "Dokl. Akad. Nauk SSSR",
	volume = "177",
	number = "1",
	pages = "70--71",
	year = "1967"
}

@article{Gubser:1998bc,
	author = "Gubser, S. S. and Klebanov, Igor R. and Polyakov, Alexander M.",
	title = "{Gauge theory correlators from noncritical string theory}",
	eprint = "hep-th/9802109",
	archivePrefix = "arXiv",
	reportNumber = "PUPT-1767",
	doi = "10.1016/S0370-2693(98)00377-3",
	journal = "Phys. Lett. B",
	volume = "428",
	pages = "105--114",
	year = "1998"
}

@article{Padmanabhan:2009vy,
	author = "Padmanabhan, T.",
	title = "{Thermodynamical Aspects of Gravity: New insights}",
	eprint = "0911.5004",
	archivePrefix = "arXiv",
	primaryClass = "gr-qc",
	doi = "10.1088/0034-4885/73/4/046901",
	journal = "Rept. Prog. Phys.",
	volume = "73",
	pages = "046901",
	year = "2010"
}

@article{Zee:1980sj,
	author = "Zee, A.",
	title = "{Spontaneously Generated Gravity}",
	reportNumber = "UPR-0154T",
	doi = "10.1103/PhysRevD.23.858",
	journal = "Phys. Rev. D",
	volume = "23",
	pages = "858",
	year = "1981"
}

@article{Adler:1980pg,
	author = "Adler, Stephen L.",
	title = "{A Formula for the Induced Gravitational Constant}",
	reportNumber = "Print-80-0439 (IAS,PRINCETON)",
	doi = "10.1016/0370-2693(80)90478-5",
	journal = "Phys. Lett. B",
	volume = "95",
	pages = "241",
	year = "1980"
}

@article{Adler:1980bx,
	author = "Adler, Stephen L.",
	title = "{Order R Vacuum Action Functional in Scalar Free Unified Theories with Spontaneous Scale Breaking}",
	reportNumber = "Print-80-0324 Rev.(IAS,PRINCETON), PRINT-80-0324 (IAS,PRINCETON)",
	doi = "10.1103/PhysRevLett.44.1567",
	journal = "Phys. Rev. Lett.",
	volume = "44",
	pages = "1567",
	year = "1980"
}

@article{Adler:1982ri,
	author = "Adler, Stephen L.",
	title = "{Einstein Gravity as a Symmetry-Breaking Effect in Quantum Field Theory}",
	reportNumber = "PRINT-81-0725-REV. (IAS,-PRINCETON), PRINT-81-0725 (IAS,-PRINCETON)",
	doi = "10.1103/RevModPhys.54.729",
	journal = "Rev. Mod. Phys.",
	volume = "54",
	pages = "729",
	year = "1982",
	note = "[Erratum: Rev.Mod.Phys. 55, 837 (1983)]"
}

@article{Zee:1978wi,
	author = "Zee, A.",
	title = "{A Broken Symmetric Theory of Gravity}",
	reportNumber = "UPR-0110",
	doi = "10.1103/PhysRevLett.42.417",
	journal = "Phys. Rev. Lett.",
	volume = "42",
	pages = "417",
	year = "1979"
}

@article{Jenkins:2009un,
	author = "Jenkins, Alejandro",
	title = "{Constraints on emergent gravity}",
	eprint = "0904.0453",
	archivePrefix = "arXiv",
	primaryClass = "gr-qc",
	reportNumber = "MIT-CTP-4025",
	doi = "10.1142/S0218271809015941",
	journal = "Int. J. Mod. Phys. D",
	volume = "18",
	pages = "2249--2255",
	year = "2009"
}

\end{document}